\documentclass[final,3p]{elsarticle}

 \usepackage{graphics}
 \usepackage{graphicx}
\usepackage{caption}
\usepackage{float}
\usepackage{subcaption}
\usepackage{amssymb}
\usepackage{amsmath}
\usepackage{color}
\usepackage{soul}           
\usepackage{ulem}
\usepackage{tikz}
\usepackage{makecell}
\usepackage{algorithm}
\usepackage{algpseudocode}
\usetikzlibrary{shapes.geometric, arrows}

\tikzstyle{normal} = [rectangle, rounded corners, minimum width=3cm, minimum height=1cm,text centered, draw=black, fill=red!30,text width=10em]

\tikzstyle{robust} = [trapezium, trapezium left angle=70, trapezium right angle=110, minimum width=3cm, minimum height=1cm, text centered, draw=black, fill=blue!30,text width=4em]

\usepackage[shortlabels]{enumitem}

\usepackage[T1]{fontenc}
\usepackage{hyperref}
\hypersetup{
colorlinks=true
}
\usepackage{epstopdf}
\usepackage{xcolor}

\biboptions{sort&compress}

\DeclareMathOperator{\cov}{Cov}
\DeclareMathOperator{\cor}{Corr}
\journal{arXiv}

\begin{document}

\begin{frontmatter}

\title{Applications of robust statistics for cyclostationarity detection in non-Gaussian
signals for local damage detection in bearings}

\author[label2]{Wojciech \.Zu{\l}awi\'nski}\corref{cor1}\ead{wojciech.zulawinski@pwr.edu.pl}
    \author[label4]{J\'{e}r\^{o}me Antoni} 
   \author[label1]{Rados{\l}aw Zimroz} 
\author[label2]{Agnieszka Wy\l oma\'nska}

\cortext[cor1]{Corresponding author.}\address[label2]{Faculty of Pure and Applied Mathematics, Hugo Steinhaus Center, Wroclaw University of Science and Technology, Hoene-Wronskiego 13c, 50-376 Wroclaw, Poland}
\address[label4]{INSA Lyon, LVA, UR677, 69621, Villeurbanne, France}
\address[label1]{Faculty of Geoengineering, Mining and Geology, Wroclaw University of Science and Technology, Na Grobli 15, 50-421 Wroclaw, Poland }

\begin{abstract}
Signals with periodic characteristics are ubiquitous in real-world applications. One of these areas is condition monitoring, where the vibration signals from rotating machines naturally display periodic behavior.  Thus, the cyclostationary analysis has evolved into the investigation of such signals. For the traditional cyclostationary approaches, the autocovariance function (ACVF) and its bi-frequency representation, spectral coherence (SC), are regarded as the base.  However, recent research has revealed that real vibration signals increasingly exhibit impulsive behavior in addition to periodicity.  As a result, there was a need for new methods to identify periodic behavior that take into account the impulsiveness of the data. In this article, we provide a way to improve the SC method by using its robust variants in place of the classical ACVF estimator (sample ACVF). The suggested concept is intuitive  and relatively simple. We create robust versions of the SC algorithm that  more accurately detect periodic behavior in signals with significant disruptions in contrast to the classical techniques. The efficiency of the proposed approach is demonstrated for simulated signals with three different types of non-Gaussian noise distribution and different levels of periodic impulses imitating a local damage. The introduced approach is also validated on real vibration signal from the  rolling element bearings operating in a crushing machine.

\end{abstract}


\end{frontmatter}

\section{Introduction}
In this paper, we discuss the problem of periodic behavior identification for random signals with non-Gaussian distributions with possible large (in absolute value) observations. Such distributions belong to the class of heavy-tailed family. The issue is presented in the context of vibration-based condition monitoring, however, it can be considered as a general problem in various applications.  Real signals frequently exhibit periodic behavior. The periodicity can occasionally be seen in the raw time series, however, it can also manifest itself in the signal's characteristics, such as functions describing its dependence structure. For second-order (i.e. with finite variance) random signals the most common functions describing the internal dependence are the autocovariance function (ACVF) or its normalized version, the autocorrelation function (ACF). Both functions  belong to the class of the so-called second-order statistics. The signal is referred to as periodically correlated or second-order cyclostationary when the ACVF and ACF are periodic in time. Cyclostationary signals are encountered in a number of actual contexts. One of the important is the condition monitoring \cite{antoni2004cyclostationary}, where the vibration or acoustic signals produced by rotating machines naturally possess periodic characteristics \cite{ZHU2005467,DELVECCHIO2018661,LAMRAOUI2014177,8553568,9388715,fe}. 

Cyclic spectral analysis, which is based on the spectral coherence (SC) statistic, is a key method used for second-order cyclostationary signals. The SC is a powerful tool for identifying the cyclic behavior of a signal in the frequency domain. It is a bi-frequency map that represents the structure of dependence (expressed by ACVF) in the time domain. As a result, the estimator of ACVF is used in a natural way in algorithms for SC computations, see e.g., \cite{antoni2007cyclic2}. The classical estimator of the ACVF, known as the sample ACVF, is utilized in traditional SC analysis.  The methods based on SC (and thus ACVF) for cyclostationary signals are widely discussed in the literature. For instance, the authors of \cite{antoni2007cyclic,csc2} used new algorithms for cyclic spectral approaches for mechanical systems, whereas the study presented in \cite{CHEN2020106683} proposed a novel method for cyclic spectrum analysis improved by machine learning techniques. Several of cyclostationary indicators (in the context of gear fault monitoring) that also make use of the ACVF were proposed in \cite{indykatory}. It is also worth mentioning other interesting approaches for cyclostationary behavior identification, such as the methodology based on the envelope spectrum  \cite{BORGHESANI201338} or approaches based on the  statistical definition of cyclostationarity \cite{PANCALDI2021107329}.

In addition to the second-order cyclostationarity analysis (based on the second-order statistics) there are known methods based on higher-order statistics for cyclostationarity detection (cyclostationarity of higher order). The second-order statistics appear to be more effective for signals with low signal-to-noise ratios, while  the higher-order analysis  (typically  more complex) when the hidden periodicity is contained in the the higher-order statistics \cite{IZZO1996303}. We refer the readers to the bibliography positions \cite{spooner1994cumulant1,spooner1994cumulant}, where the authors applied the cumulant theory for higher-order cyclostationary signals, see also \cite{higher_order11,higher_order2,higher_order12} and references therein.

Although the classical theory of cyclostationary processes has evolved over time and new dedicated methods are still being proposed \cite{BOUDOU2022104875}, in real applications we are increasingly observing behavior indicating heavy-tailed distribution of the signals manifested by their impulsive behavior. The condition monitoring literature has discussed this issue in various contexts for many years. 
Note that the impulsive behavior of real signals may be related to particular machine operations (cutting, crushing, milling, drilling, compression, etc.) \cite{Kuba10,nowicki2020local, borghesani2017cs2,wodecki2019impulsive, Lahrache201721,mssp1} or to externally generated, completely random disturbances as well as noise that appears during data transmission or even numerical issues during data processing \cite{Mauricio2020}.  In recent studies the problem of  heavy-tailed distributed signals was also discussed by many authors in the condition monitoring context, see e.g. \cite{MA2021108983,PENG2021107786,SKOWRONEK2023110465}. It should be highlighted that if the background noise is non-Gaussian heavy-tailed distributed, then classical cyclostationary analysis  may be of small efficiency and the known techniques may not return the expected information. This problem is raised for instance in  \cite{BARSZCZ2011431,BARSZCZ20091352,URBANEK20121782,URBANEK201713}. In the articles  \cite{zak2017measures,Schmidt2021,Wodecki2021,kruczek2020detect,borghesani2017cs2} the authors tested the efficiency of the classical algorithms  for heavy-tailed distributed signals and demonstrated their limitations for identifying cyclostationary behavior. 
As a result, in the latest research, a number of improvements have been proposed to the traditional cyclostationarity detection techniques for signals with heavy-tailed distributions. Among the many modern approaches, we  mention here the articles \cite{BERSHAD2019149,SHLEZINGER2019260}, where authors studied the stochastic behavior of the least mean squares algorithm in a system identification framework for cyclostationary signals without assuming their Gaussian distribution; an approach based on the high-dimensional space transformation \cite{9514478}; methods based on the fractional lower-order statistics \cite{6040125,zak1} or algorithms utilizing the cyclic correntropy (and its spectral version) \cite{cyclic_corre2,cyclic_corre1,cyclic_corre3}. We highlight that Napolitano in \cite{NAPOLITANO2016385} also raised the problem of identification of cyclic behavior of heavy-tailed distributed non-Gaussian signals as one of the crucial issues for real applications.

In recent years, one can also observe the development of techniques dedicated to impulsive cyclostationarity signals in condition monitoring area. For instance in \cite{cyclic_corre4} Zhao et al. introduced a cyclic correntropy for rolling element bearing fault diagnosis when the signal has impulsive behavior. In 
 \cite{mssp1} the new statistical detection method for rolling element bearing faults based on $\alpha$-stable distribution was discussed. In \cite{hebda2022infogram} the improved infogram was introduced for local damage detection while in \cite{cvb} the conditional variance-based statistic was proposed as the selector for informative frequency band identification. In \cite{PENG2023110351} the authors proposed to apply generalized Gaussian cyclostationarity for blind deconvolution for  bearing fault diagnosis under non-Gaussian conditions. Additionally, in \cite{ANTONI2019290} a statistical methodology based on the maximum likelihood ratio as a general framework to design condition indicators for cyclostationary non-Gaussian signals was discussed, while in \cite{HE2021108738} a new strategy for multi-fault detection of gearboxes with  impulsive fault components was introduced. 

In this study, we also extend the classical methodology of second-order cyclostationary analysis for signals that may have a non-Gaussian heavy-tailed distribution. Although the methodology is shown in the setting of local damage detection, it can be viewed as universal and used to the detection of cyclostationarity in various contexts.  The idea is intuitive and relatively simple. In the classical algorithm for SC calculation we propose to switch out the traditional ACVF (or ACF) estimator with more robust versions that are less susceptible to strong impulses. When a heavy-tailed distribution of the underlying signal is anticipated, robust statistics are frequently applied, see e.g., \cite{cvb} where the robust estimator for variance is discussed. 
In the literature there are few known efficient ACVF (and ACF) estimators that are less sensitive to large observations, while still providing a good approximation to the classical measure, see e.g. \cite{CHAN1992149,pz2,pz3,pz4,pz5,kend3}. Here, we outline the overall methodology for enhancing the SC algorithm through the use of robust ACVF (or ACF) estimators. In addition to the three robust statistics for which the analyses are presented, the procedure can be used with any other robust estimators of the classical measure. Let us point out that the notion of making the SC statistic more effective is not new, and certain ideas have been made in the literature. In   \cite{540601} the authors proposed to apply the robust estimation of cyclic correlation in contaminated Gaussian noise while in \cite{5200382} the robust non-parametric cyclic correlation-based spectrum was proposed. Chen et al. in \cite{CHEN2020106683} introduced a hybrid method for bearing fault diagnosis that enhances the cyclic SC (CSC) by deep learning algorithm. Li et al. proposed  an enhanced rolling bearing fault detection method that combines the SC with  sparse code shrinkage denoising \cite{LI2020335}. Last but not least, in our previous research \cite{kruczek2021generalized} we explored generalized spectral coherence, where the measure of dependence designated for the special case of the non-Gaussian signal ($\alpha$-stable distributed) replaced the ACVF in the SC formulation, see also \cite{kruczek2020detect,nowicki2020local,electronics10151863} for other dependence measures designed for signals with specific non-Gaussian  distributions.

On the one hand, the universality of our proposition is related to the overall idea of substituting the classical estimator of the ACVF (or ACF)  by  robust statistics in the SC algorithm. On the other hand, the universality can be attributed to the fact that utilizing robust measures does not require the signal to have a particular distribution, as it did in \cite{kruczek2021generalized}, for example. The presented simulation studies provided for three different types of heavy-tailed distributed noise  clearly confirm this thesis and efficiency of the proposed methodology in contrast to the classical approach. The simulation studies are supported by real data analysis, where a vibration signal with heavy-tailed distribution is examined in the context of the presented methodology.

The rest of the paper is organized as follows. In Section \ref{sec_1}
we present the signal model and discuss three distributions which characterize the noise in the model under consideration.  In Section \ref{sec3} we present three selected robust indicators of periodic behavior applied for the analysis in the bi-frequency domain. The general methodology is based on the robust estimators of ACVF (or ACF) that are utilized in algorithm for SC calculation.   In Section \ref{sec:sim}
we present a simulation study for three different types of the noise. The efficiency of the SC algorithm based the robust statistics is demonstrated for various levels of non-Gaussianity of the used distributions. The comparative study with the classical (based on sample ACVF) and generalized (based on the measure dedicated to $\alpha$-stable noise) approaches is provided.   In Section \ref{real} an analysis of a real signal from a mining machine is  presented. Here, we use the real signal from the healthy machine and add artificial damage of various amplitudes in order to control the level of amplitude of cyclic impulses (related to damage). The same approach was proposed in \cite{kruczek2021generalized}. The presented results  illustrate the effectiveness of the suggested robust approaches in comparison to the existing methodologies also for real cases. The last section concludes the paper.

\section{Model of the signal}\label{sec_1}

We assume the following model of the signal $\{X_t\}$, 
\begin{eqnarray}\label{model2}
X_t=s(t)+Z_t, \quad t\in\mathbb{Z}
\end{eqnarray}
where $s(t)$ is a deterministic component, called the signal of interest (SOI), and $\{Z_t\}$ is a random component -- a sequence of i.i.d. random variables with zero mean -- called noise (note the difference in notation intended to recognize a deterministic component from a random component). Here, we assume a specific form of $s(t)$, which is composed of a series of individual impulses located in time with a given period $T=1/f_f$, where $f_f$ is a  fault frequency.
A single impulse may be specified as a decaying harmonic oscillation of the following form
\begin{equation} \label{imp_structure}
    h(t)=B\sin{(2\pi f_c t)}e^{-dt}, \quad t\geq 0
\end{equation}
$B$ is the amplitude, $f_c$ is the carrier frequency (related to the structural resonance in the machine), $d$ is the decay coefficient of the exponential function, and $t$ represents time.

{The signal given in Eq. (\ref{model2}) with $\{Z_t\}$ being a Gaussian i.i.d. sequence is a simplified model of the  signal from faulty bearings \cite{rz1,rz2,rz3,rz4,rz5}. In this case, the components of the vibration signal (\ref{model2}) are independent sources and could be represented as 
random Gaussian component $\{Z_t\}$ related to bearings in healthy state and the SOI to the  impulsive cyclic component related to local damage.}  However, when $\{Z_t\}$ is non-Gaussian heavy-tailed distributed i.i.d., then it is considered as a noise with large impulses. This kind of model was considered in \cite{wylomanska2016impulsive,kruczek2020detect,kruczek2021generalized}, where the periodic component was disturbed by   non-Gaussian noise.

\subsection{Selected distributions of the noise }\label{ANdists}
In order to demonstrate the universality of the methodology introduced in this paper,  we consider three types of noise distributions $\{Z_t\}$ in the model (\ref{model2}). 

The first case we assume that the noise constitutes a sequence of i.i.d. random variables and  for each $t\in \mathbb{Z}$, $Z_t$ is defined as follows
\begin{eqnarray}\label{mG}
Z_t=\xi_t+A_t K_t,
\end{eqnarray}
 where $\{K_t\}$ is an i.i.d. sequence of the following distribution
\begin{eqnarray}\mathbb{P}(K_t=1)=\mathbb{P}(K_t=-1)=p/2, ~~\mathbb{P}(K_t=0)=1-p,~~p\in[0,1]\end{eqnarray}
and $\{\xi_t\}$ is i.i.d.   Gaussian time series with zero mean and variance $D^2$. Moreover, $A_t$ also constitutes an i.i.d. sequence. We assume $A_t$, $K_t$ and $\xi_t$ are independent for each $t$. Let us note that for $p=0$ the sequence $\{Z_t\}$ is just Gaussian distributed white noise. One may assume that $A_t$ is a constant or that it is a random variable from a given distribution. In the case when $A_t$ is fixed, the random variable defined in  Eq. (\ref{mG}) has the so-called mixture of Gaussian distribution. In this paper, we consider the case when $A_t$ is a uniformly distributed random variable $A_t\sim U(0,a)$ ($a>0$) and we take the notation $\mathcal{M}(a,p,D)$. Let us note that the random variable from  $\mathcal{M}(a,p,D)$ distribution has finite variance for any set of parameters.

As the second  distribution of the  noise  in model (\ref{model2}) we consider the scaled Student's t  $\mathcal{T}(\nu,\delta)$. More precisely, we assume that  $\{Z_t\}$ is a sequence of i.i.d. random variables and for each $t\in \mathbb{Z}$, $Z_t$ has scaled Student's t distribution defined through the probability density function (PDF) \cite{student1,student2}
        \begin{equation}
    f(z) = \frac{\Gamma\left( \frac{\nu+1}{2}\right)}{\delta \sqrt{\nu \pi} \Gamma\left( \frac{\nu}{2}\right)} \left[ \frac{\nu + \frac{z^2}{\delta^2}}{\nu}\right]^{-\left(  \frac{\nu+1}{2}\right)},\quad\quad z\in \mathbb{R},
    \end{equation}
where $\Gamma(\cdot)$ is the gamma function, $\nu >0$ is a  shape parameter and $\delta >0$ is a scale parameter. The variance for this distribution is defined only for $\nu >2$. Otherwise, it is infinite.   When $\nu$ tends to infinity, the scaled Student's t distribution tends to a Gaussian distribution and thus for large $\nu$ the sequence $\{Z_t\}$ tends to i.i.d. Gaussian distributed noise.

In the last case we assume that the noise $\{Z_t\}$ in model (\ref{model2}) is a sequence of i.i.d. random variables with an $\alpha$-stable distribution (denoted further as $\mathcal{S}(\alpha,\sigma)$). This distribution has been successfully used in condition monitoring applications by various authors, see e.g. \cite{mssp1,zak3,zak56,8726382,7819828,6911199,6324624}.  In this paper, we consider the symmetric version of the $\alpha$-stable distribution defined through the characteristic function \cite{stable1,stable2}
\begin{eqnarray}
  \Phi(z)=\mathbb{E}\exp\left\{iZ_tz\right\}=\exp{\left( -\sigma^{\alpha} |z|^{\alpha}\right)}, \quad \quad z \in \mathbb{R}
\end{eqnarray}
 where $0<\alpha\leq2$ is the stability index and $\sigma>0$ is the scale parameter. The symmetric $\alpha$-stable distribution has no closed-form PDF and cumulative distribution function (CDF). The only exception is the Gaussian distribution (that is, for $\alpha=2$) and the Cauchy distribution (that is, for $\alpha=1$). The stability index is responsible for the heaviness of this distribution's tail, $1-F(z)\sim z^{-\alpha}$ ($F(\cdot)$ is a CDF of random variable $Z_t$), i.e. the smaller $\alpha$, the probability of large values is  much higher. For $\alpha<2$, the variance  of $\alpha$-stable distribution is infinite.

The above three distributions belong to the wide class of heavy-tailed distributions (except the Gaussian cases), however their nature is different.  All of them are continuous distributions, but $\mathcal{M}(a,p,D)$ and  $\mathcal{T}(\nu,\delta)$  for $\nu>2$ are finite-variance distributions, while in the  $\alpha$-stable case only for $\alpha=2$ the variance is finite.

\section{Robust indicators of periodic behavior}\label{sec3}
The basic characteristics used for the periodic behavior identification for finite-variance signals are the autocovariance function and the autocorrelation function. In this paper, we also take into account this approach and analyze the periodic behavior indicators based on the ACVF/ACF  in the bi-frequency domain. However, it should be mentioned there also exist other measures that may indicate the presence of a periodic (or cyclostationary) behavior, see e.g. \cite{kruczek2020detect,kruczek2021generalized}. 

Let us emphasize, the ACVF (or ACF) of given time series for time points $t$ and $t+h$ ($t,h\in \mathbb{Z}$)  can be considered as the classical covariance (or correlation) between two random variables, where the first represents the analyzed time series at time point $t$ while the second one at time $t+h$. In Section \ref{robust_time} we present the selected robust estimators of correlation between two random variables based on vectors of observations corresponding to them. In Section \ref{CSCsubsec} we present how to use them to obtain a robust version of CSC map.

Let us assume $W_1$ and $W_2$ are two finite-variance random variables with zero means. The theoretical covariance between them is a measure defined as
\begin{eqnarray}\label{measure0}
\cov(W_1,W_2)=\mathbb{E}W_1W_2
\end{eqnarray}
while the theoretical correlation (called Pearson correlation) is given by
\begin{eqnarray}\label{measure}
\cor(W_1,W_2)=\frac{\cov(W_1,W_2)}{\sqrt{\text{Var}(W_1)\text{Var}(W_2)}},
\end{eqnarray}
where $\text{Var}(W_j)$ is the variance of random variable $W_j$, $j=1,2$. As mentioned, when the random variables $W_1$ and $W_2$ correspond to the same process but in different time points, then the measure defined in Eq. (\ref{measure}) is called the autocorrelation function. Similarly, in that case the measure given in Eq. (\ref{measure0}) is called the autocovariance function. The classical estimator of the measure given in Eq. (\ref{measure}) is the sample correlation; for the zero-mean vectors $\mathbf{w}_1=(w^1_1,w^2_1,\cdots,w^N_1)$ and $\mathbf{w}_2=(w^1_2,w^2_2,\cdots,w^N_2)$ of length $N$ corresponding to zero-mean random variables $W_1$ and $W_2$, respectively, it is given by 
\begin{eqnarray}\label{measure1}
M_1(\mathbf{w}_1,\mathbf{w}_2)=\frac{M_1'(\mathbf{w}_1,\mathbf{w}_2)}{\frac{1}{N}\sqrt{\sum_{i=1}^N (w^i_1)^2\sum_{i=1}^N (w^i_2)^2}},
\end{eqnarray}
where
\begin{eqnarray}\label{measure1_prime}
  M_1'(\mathbf{w}_1,\mathbf{w}_2) =\frac{1}{N} \sum_{i=1}^N w^i_1 w^i_2.
\end{eqnarray}
The statistic $M_1$ is called the sample correlation. In case when the vectors $\mathbf{w}_1$ and $\mathbf{w}_2$ are realizations of the same process but in different time points, then the statistic given in Eq. (\ref{measure1}) is called the sample autocorrelation function (sample ACF), while $M_1'$ is called the sample autocovariance function (sample ACVF). 

The statistics given in Eqs. (\ref{measure1_prime}) and (\ref{measure1}) are considered as  efficient estimators of the measures (\ref{measure0}) and (\ref{measure}), respectively, in the classical case (i.e. forfinite-variance distributed random variables) for the identification of linear dependence between the vectors $\mathbf{w}_1$ and $\mathbf{w}_2$. However, their main disadvantage  is related to their sensitivity to outliers that may appear in the vectors of observations. In that case, the information about the dependence between the random variables $W_1$ and $W_2$ may be disturbed when the classical estimator is applied. This point was discussed e.g. in {\cite{kruczek2020detect,kruczek2021generalized}}, where the classical estimator defined in (\ref{measure1}) was confronted with the estimator of the  measures dedicated for $\alpha$-stable distributed time series in case of non-Gaussian signals. Thus, in the literature, there are considered substitutes of $M_1$ (or $M_1'$) that also estimate the theoretical correlation given in Eq. (\ref{measure}), but are insensitive to large impulses visible in the data.  In this paper, we present  three selected robust estimators of the classical measures, however the proposed methodology is universal and any other robust sample measure can be used for the considered problem. 

\subsection{Robust estimators of correlation}\label{robust_time}
The first considered robust estimator of the correlation (\ref{measure}) is called trimmed. It is based on  the estimator $M_1$ (see Eq. (\ref{measure1})), but it ignores the most extreme observations. First, based on vectors $\mathbf{w}_1$ and $\mathbf{w}_2$ we construct the vector
\[\mathbf{w}_3=(w^1_3,w^2_3,\cdots,w^N_3)=(w^1_1w^1_2,w^2_1w^2_2,\cdots,w^N_1w^N_2).\]
Let $c$ be a trimming constant, i.e. "fraction" of data in the sample $\mathbf{w}_3$ of the predicted number of outliers.
Then, we define trimmed vectors $\tilde{\mathbf{w}}_1$ and $\tilde{\mathbf{w}}_2$ in the following way 
\[\tilde{\mathbf{w}}_k = \{w_k^i\: : \:i=1,\ldots,N;\;w_3^{(g)}<w^i_3<w_3^{(n-g+1)}\},\quad k=1,2,\]
where $g=\lfloor c\cdot N\rfloor$ for a trimming constant $0\leq c<0.5$, and $w_3^{(1)}, w_3^{(2)},\dots, w_3^{(N)}$ is an ordered vector $\mathbf{w}_3$ in ascending order. Finally, we calculate the trimmed estimator
\begin{equation}
    M_2^c(\mathbf{w}_1,\mathbf{w}_2)=M_1(\tilde{\mathbf{w}}_1,\tilde{\mathbf{w}}_2).
\end{equation}
We recall, the choice of $c$ is a crucial issue, as higher values of $c$ can increase robustness but decrease the efficiency for samples without outliers, meanwhile lower values might not trim out enough extreme values. In \cite{CHAN1992149}, authors recommend using $c=3-5\%$ for medium contaminated series, however, there are also considered other versions of the trimmed statistic. 

The second robust estimator of correlation (\ref{measure}) discussed in this paper is based on the Kendall correlation \cite{kendall1938new}, which for the vectors $\mathbf{w}_1, \mathbf{w}_2$ is defined as 
   \begin{eqnarray}
   \rho_K(\mathbf{w}_1,\mathbf{w}_2) =\frac{ 2 }{ N(N-1) } \sum_{1 \leq i \leq j \leq N} J(i,j),
   \label{measure3}
\end{eqnarray}
where $J(i,j) =\text{sgn}(w_1^i-w_2^i) \text{sgn}(w_1^j-w_2^j)$ and $J(i,j)= 1,$ if a pair $(w^i_1,w^i_2)$ is concordant with a pair $(w^j_1,w^j_2),$ i.e. if $(w^i_1-w^j_1)(w^i_2-w^j_2) > 0; J(i,j) =-1,$ if a pair $(w^i_1,w^i_2)$ is discordant with a pair $(w^j_1,w^j_2),$ i.e. if $(w_1^i-w^j_1)(w^i_2-w^j_2) < 0$. $\text{sgn}(\cdot)$ is a signum function.

The Kendall correlation coefficient is based on the difference between the probability that two variables are in the same order (for the observed data vector) and the probability that their order is different. The formula for Kendall correlation (\ref{measure3}) requires that the variable values can be ordered. This factor takes values in $[-1,1]$. The value equal to $1$ means the full accordance, the value equal to $0$ indicates no relation between the analyzed vectors, while the value equal to $-1$ means the complete opposite accordance. The Kendall correlation indicates not only the strength but also the direction of the dependence. Similar to the trimmed statistic, it is resistant to outliers. The Kendall correlation-based estimator of the Pearson correlation is given by \cite{pz4}
\begin{eqnarray}\label{kendallm3}
    M_3(\mathbf{w}_1,\mathbf{w}_2) = \sin \left(\frac{\pi \rho_K(\mathbf{w}_1,\mathbf{w}_2)}{2}\right).
\end{eqnarray}
The above relation holds for elliptically symmetric distributions (e.g. multivariate Gaussian one).

The last considered robust estimator for (\ref{measure}) is based on the Spearman correlation that belongs to the family of rank estimators. The Spearman correlation for vectors $\mathbf{w}_1, \mathbf{w}_2$ is defined as \cite{kend3}
\begin{equation}\label{eq:spearman_1}
   \rho_S(\mathbf{w}_1,\mathbf{w}_2)= \frac{\sum_{i=1}^{N}r_1^i r_2^i}{\sqrt{\sum_{i=1}^{N}(r_1^i)^2 \sum_{i=1}^{N}(r_2^i)^2}},
\end{equation}
where $\mathbf{r}_i$, $i=1,2$ is the zero-mean vector of ranks corresponding to $\mathbf{w}_i$. In case where there are no tied ranks (i.e., $\mathbf{r}_i$ contains only unique values), one can also calculate the Spearman correlation in the following way:
\begin{equation}\label{eq:spearman}
   \rho_S(\mathbf{w}_1,\mathbf{w}_2)= 1 - \frac{6\sum_{i=1}^N (r_1^i - r_2^i)^2 }{N(N^2-1)}.
\end{equation}
From \eqref{eq:spearman_1} we see that  the Spearman correlation is equivalent to the Pearson correlation between vectors of ranks, i.e. $\rho_S(\mathbf{w}_1,\mathbf{w}_2)=M_1(\mathbf{r}_1,\mathbf{r}_2)$. It takes values in $[-1, 1]$ and tests a monotonic relationship between two vectors. The Spearman correlation-based estimator of the Pearson correlation is given by \cite{pz4}:
\begin{eqnarray}\label{spearmanm4}
    M_4(\mathbf{w}_1,\mathbf{w}_2) = \sin \left(\frac{2\pi \rho_S(\mathbf{w}_1,\mathbf{w}_2)}{6}\right).
\end{eqnarray}
As in the case of the Kendall correlation, this relation holds for elliptically symmetric distributions.

\subsection{Robust indicators of periodicity in bi-frequency domain}\label{CSCsubsec}
In this part, we demonstrate how to apply the sample robust correlations described in Section \ref{robust_time} to the identification of the periodic behavior in the bi-frequency domain. The classical tool used to analyze the periodicity for the classical version of the model (\ref{model2}) in the case of finite-variance distributed noise $\{Z_t\}$ is the cyclic spectral coherence. This bi-frequency statistic is a double Fourier transform of the instantaneous autocovariance function, defined as follows \cite{antoni2007cyclic2}
\begin{eqnarray}\label{csc}
    |\gamma_X(f,\epsilon)|^2 = \frac{\left|S_X^{}(f,\epsilon)\right|^2}{S_X^{}(f+\epsilon/2,0)S_X^{}(f-\epsilon/2,0)},
\end{eqnarray}
where
\begin{eqnarray}\label{spect_corr}
    S_X^{}(f,\epsilon) = \lim_{N\rightarrow\infty}\frac{1}{N} \sum_{t=-N}^N \sum_{\tau=-\infty}^{\infty} R_X(t,\tau)e^{-i2\pi f\tau}e^{-i2\pi \epsilon t}
\end{eqnarray}
for $R_X(t,\tau) = \mathbb{E}X_tX_{t-\tau}$ being the autocovariance function of process $\{X_t\}$, and $\epsilon$ is the cycle frequency.

In practice, there are several ways to estimate the spectral coherence, e.g. the averaged cyclic periodogram (ACP) method \cite{antoni2007cyclic2}. As a result, we obtain a bi-frequency map. Let us note that the classical approaches for SC estimation are based on the autocovariance measure with its standard estimator defined in Eq. (\ref{measure1}). Hence, they might not perform well in case of heavy-tailed behavior of the signal. Such tendency was shown in \cite{kruczek2021generalized} for infinite-variance $\alpha$-stable distributed signals. As an alternative, the authors proposed there the generalized spectral coherence (GSC), where the autocovariance is replaced by the autocovariation measure. In consequence in the algorithm for SC calculation, the sample ACVF is replaced by the estimator  of the autocovariation. In this paper, we consider different approaches. Although we still consider the autocovariance-based SC statistic, we use a robust ACVF/ACF estimators. Similarly as in \cite{kruczek2021generalized}, for SC calculation we apply the ACP algorithm. The considered procedure is presented in Algorithm \ref{algo:csc}. Let us note that the only change with respect to the original algorithm is in line 22 of Algorithm \ref{algo:csc}, where a robust estimator is utilized. In the original version of the algorithm, the sample ACVF is used and the final SC statistic is normalized. For trimmed estimators, we use their sample ACVF version. The Kendall and Spearman estimators are devoted only to the ACF measure. In these cases, we replace the sample ACF with a given robust estimator.

Let us also note that in Algorithm \ref{algo:csc} the value of a given estimator $M(\cdot,\cdot)$ is calculated from inputs which are complex numbers. Here, we calculate the introduced robust estimators in the following way.  Let us note that for two complex-valued random variables $\xi_1 = \Re_1 + \Im_1 j$ and $\xi_2 = \Re_2 + \Im_2 j$ we have
\begin{equation}
     \mathbb{E}\xi_1\xi_2^* = \mathbb{E}(\Re_1\Re_2) - \mathbb{E}(\Im_1\Im_2) + [\mathbb{E}(\Re_1\Im_2)+\mathbb{E}(\Im_1\Re_2)]j,
\end{equation}
so that we obtain four real expectations where each one can be estimated by the trimmed estimator of the ACVF. In the Kendall method, the estimator is calculated in the same way as defined in \eqref{measure3} and \eqref{kendallm3}, using $\text{sgn}(z) = z/|z|$ for $z\in\mathbb{C}$. For the Spearman estimator, the proposed approach for complex numbers is the following -- we set 
\begin{eqnarray}
\rho_S(\mathbf{w}_1,\mathbf{w}_2) = M_1(\mathbf{r}_{\text{Re}(\mathbf{w_1})}+\mathbf{r}_{\text{Im}(\mathbf{w_1})}j,\mathbf{r}_{\text{Re}(\mathbf{w_2})}+\mathbf{r}_{\text{Im}(\mathbf{w_2})}j),
\end{eqnarray}
where $\mathbf{r}_{\text{Re}(\mathbf{w_i})}$ and $\mathbf{r}_{\text{Im}(\mathbf{w_i})}$ are zero-mean vectors of ranks of respectively real and imaginary parts of vector $w_i$, $i=1,2$. The Spearman-based estimator of the Pearson correlation is then calculated as before using \eqref{spearmanm4}.

In the procedure presented in Algorithm \ref{algo:csc}, we assume that the condition $f_s/f_f > n$ is met, where $f_s$ is the sampling frequency and $n$ is the length of the window function $w(\cdot)$ (see line 2).

\begin{algorithm}
  \caption{Robust spectral coherence for a signal $X = \{x_1,\ldots,x_L\} \in \mathbb{R}^L$.}\label{algo:csc}
  \begin{algorithmic}[1]
  \item[]
    \begin{enumerate}[itemsep=0pt,parsep=0pt,topsep=0pt,label=$\bullet$]
    \item $\odot$ - element-wise multiplication of vectors
    \item $X[\text{index}] = [X_{\text{index}_1},\ldots,X_{\text{index}_n}]$, where $\text{index} = [\text{index}_1,\ldots,\text{index}_n]$
    \end{enumerate}
    \State Set $M(\cdot,\cdot)$ - selected robust covariance/correlation estimator
    \State Set $w(\cdot)$ - window function of length $n$
    \State Set nfft - number of sampling points to calculate DFT
    \State Set nover - size of overlap
    \State Set $\epsilon_{\min},\,\epsilon_{\max}$ -- minimal and maximal modulation frequency
    \State $t = [0,1,\ldots,N-1]$
    \For{$k \leftarrow \epsilon_{\min} \text{ to }  \epsilon_{\max}$}
    \State $K = \left\lfloor \frac{N-\text{nover}}{n-\text{nover}} \right\rfloor$
    \State $X^k = X \odot e^{i\pi k t}$
    \State $Y^k = X \odot e^{-i \pi k t}$
    \State $\text{index} = [1,\ldots,n]$
    \For{$i \leftarrow 1 \text{ to } K$}
    \State $X^w = w \odot X^k[\text{index}]$
    \State $Y^w = w \odot Y^k[\text{index}]$
    \For{$j \leftarrow 1 \text{ to } \text{nfft}$}
    \State $X_w(j,i) = \text{DTFT}_{\text{nfft}}(j,X^w)$
    \State $Y_w(j,i) = \text{DTFT}_{\text{nfft}}(j,Y^w)$
    \EndFor
    \State $\text{index} = \text{index} + (n-\text{nover})$
    \EndFor
    \For{$j \leftarrow 1 \text{ to } \text{nfft}$}
        \State $S_X(f_j,\epsilon_k) = M(Y_w(j,:),X_w(j,:)^*)$
    \EndFor
    \State Calculate robust spectral coherence $|\gamma_X(f_j,\epsilon_k)|^2$
    \EndFor
  \end{algorithmic}
\end{algorithm}

\section{Simulated signals analysis}\label{sec:sim}

\begin{figure}
    \centering
    \includegraphics[width=0.4\textwidth]{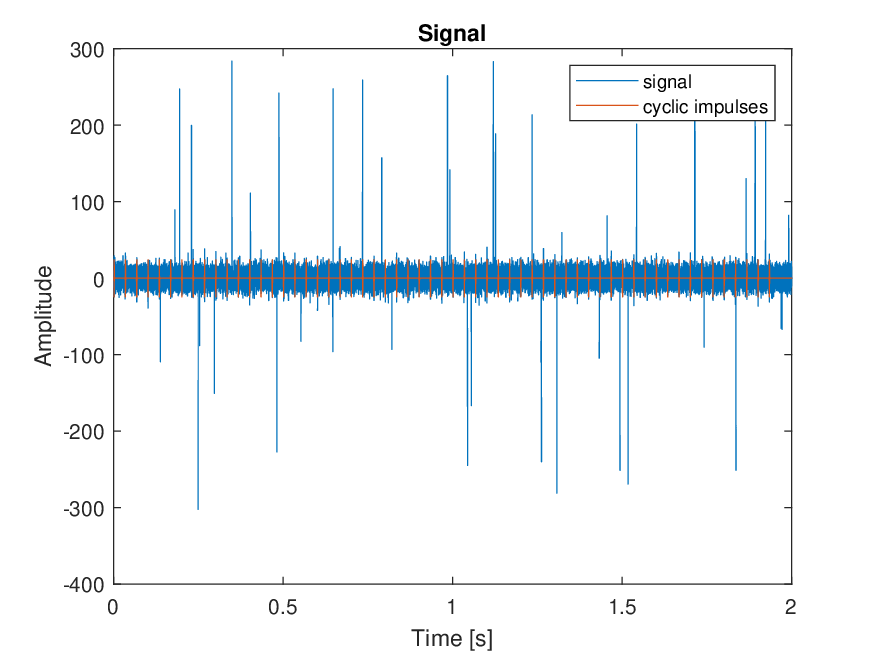}
     \includegraphics[width=0.4\textwidth]{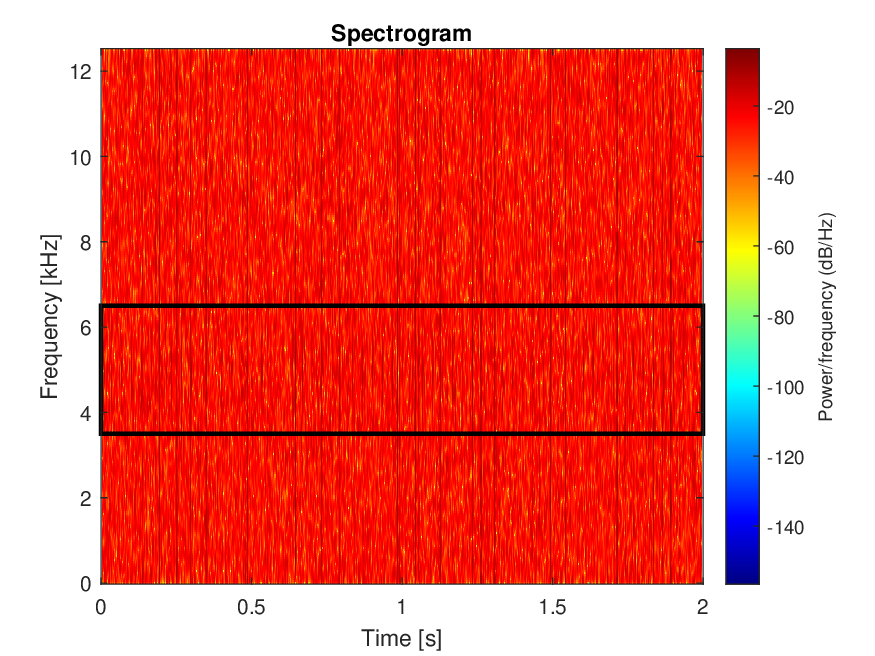}
    \caption{Sample realization of the model $\{X_t\}$ for $\{Z_t\} \sim \mathcal{M}(300,0.001,8)$ (Case 1) and its spectrogram (with the true informative frequency band marked).}
    \label{signal_M}
\end{figure}

\begin{figure}
    \centering
    \includegraphics[width=0.4\textwidth]{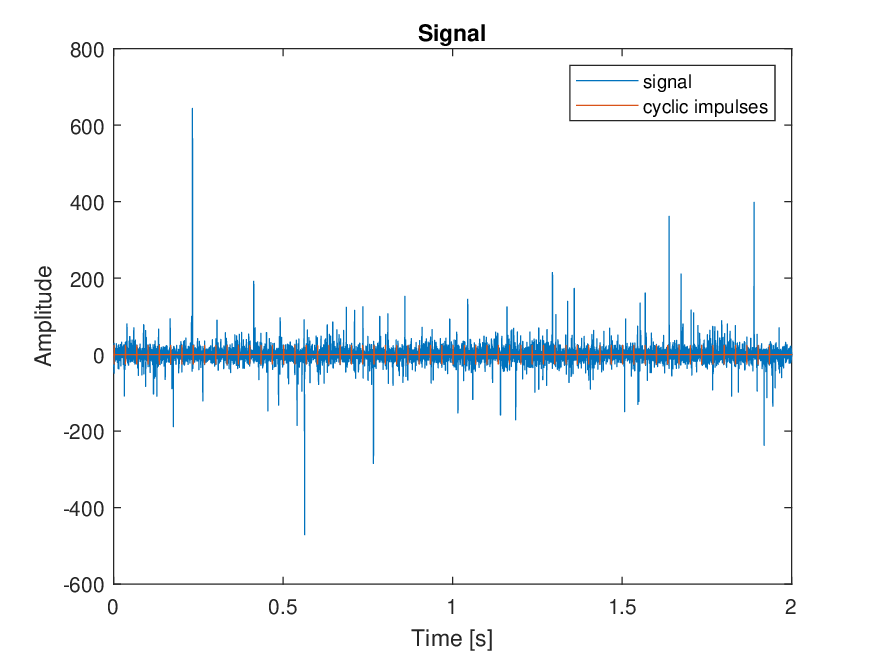}
     \includegraphics[width=0.4\textwidth]{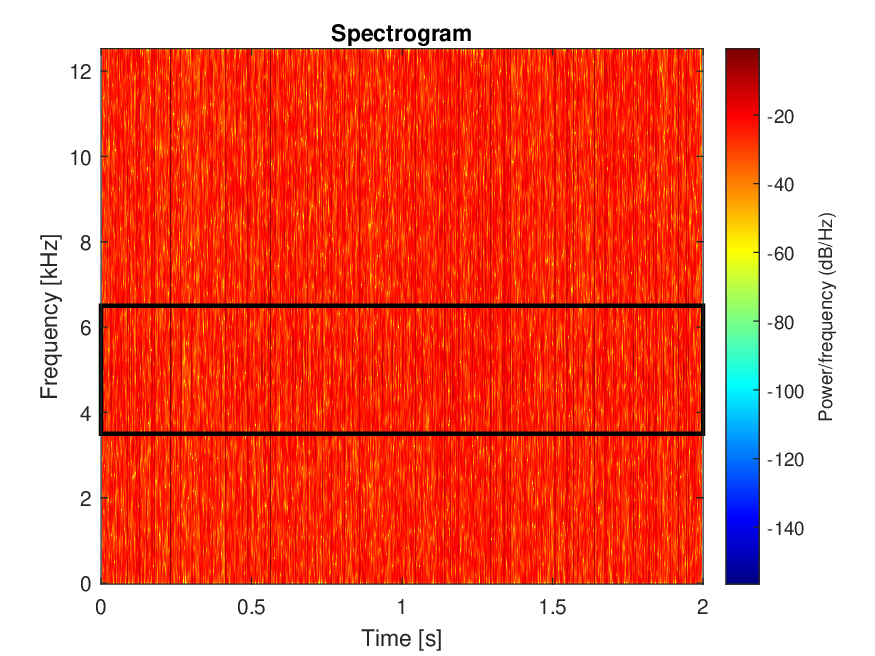}
    \caption{Sample realization of the model $\{X_t\}$ for $\{Z_t\} \sim \mathcal{T}(2,3)$ (Case 2) and its spectrogram (with the true informative frequency band marked).}
    \label{signal_t}
\end{figure}

\begin{figure}
    \centering
    \includegraphics[width=0.4\textwidth]{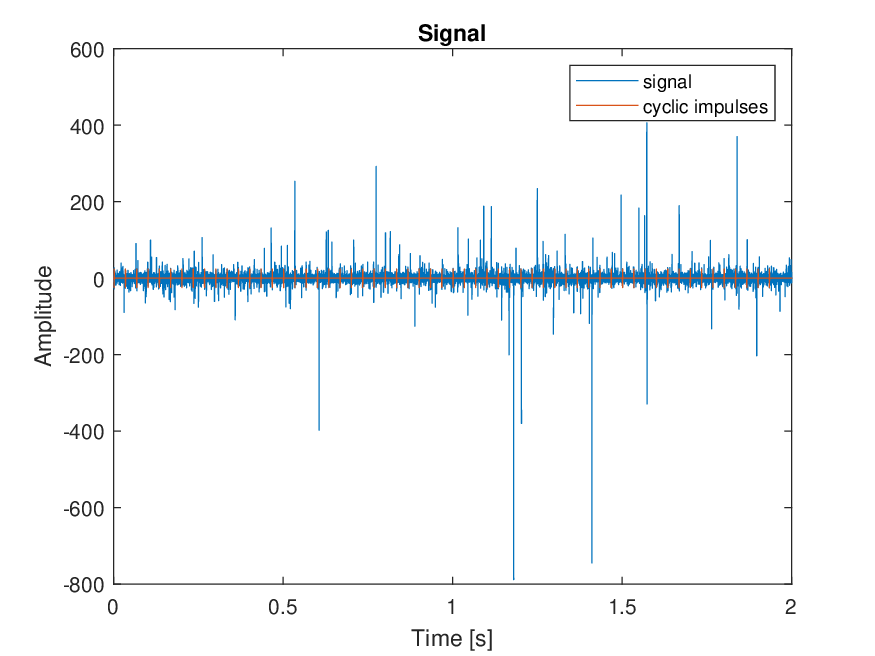}
     \includegraphics[width=0.4\textwidth]{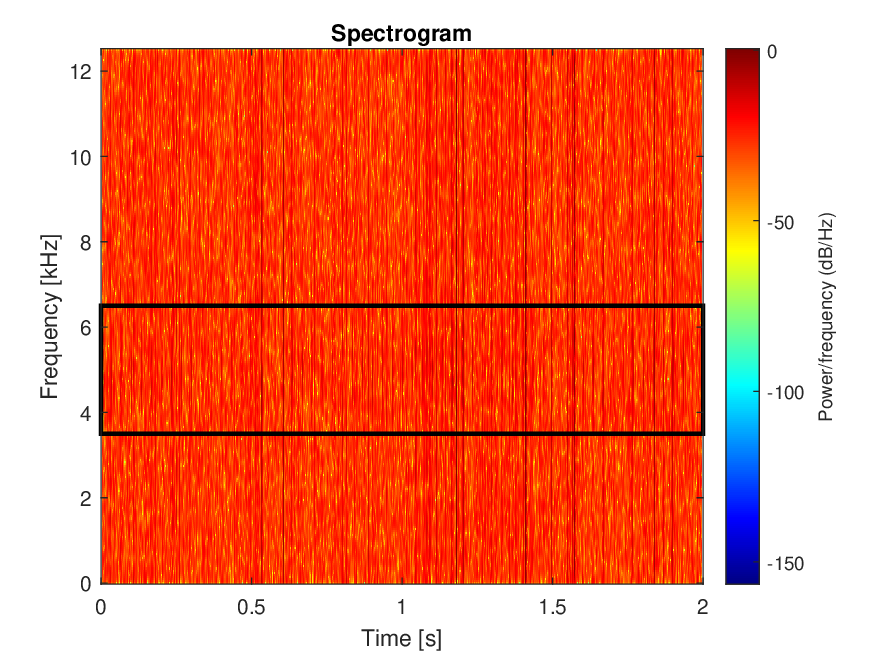}
    \caption{Sample realization of the model $\{X_t\}$ for $\{Z_t\} \sim \mathcal{S}(1.7,3)$ (Case 3) and its spectrogram (with the true informative frequency band marked).}
    \label{signal_stab}
\end{figure}

In this section, the proposed methodology is assessed and compared with the classical approach. This analysis is based on simulated signals from model \eqref{model2}. Throughout the study, we assume that the signal of interest $s(t)$ has a form of cyclic impulses with cyclic frequency $f_f = 30$ Hz,  informative frequency band $f_c = [3500, 6500]$ Hz, amplitude $B=45$, decay coefficient $d=3000$ and sampling frequency $f_s = 25000$ Hz. Moreover, we consider three distributions of the random component $\{Z_t\}$ presented in Section \ref{ANdists}: $\mathcal{M}(a,p,D)$, $\mathcal{T}(\nu,\delta)$ and $\mathcal{S}(\alpha,\sigma)$. In the following simulation study, we analyze samples of length $L=50000$ observations, i.e. 2 seconds.  In Figs. \ref{signal_M}-\ref{signal_stab}, we present sample realizations and their spectrograms for the considered model $\{X_t\}$ with $\{Z_t\}$ from $\mathcal{M}(300,0.001,8)$, $\mathcal{T}(2,3)$ and $\mathcal{S}(1.7,3)$. In this paper, to construct all spectrograms (for both simulated and real signals), we use 512 frequency points (nfft) and Hann window of length 128 with overlap 110.  
The presented signals will also be considered in the further analysis -- we will refer to them as Cases 1, 2, 3, respectively. In all of them, one can see that the signal of interest is covered by non-cyclic noise -- in particular, by the presence of outlying values. This also implies the behavior of spectrograms which do not exhibit the periodicity in a clear way.

Let us note that Cases 2 and 3 are infinite-variance signals; hence, from the theoretical point of view, the autocovariance and autocorrelation are not defined. However, as it will be presented, robust methods are still useful there despite this fact. In the following comparisons, aside from the presented sample and robust ACVF/ACF estimators, we also include the approach based on the normalized covariation, denoted further as NCV. It is an alternative measure of dependence defined for two $\mathcal{S}(\alpha,\sigma)$-distributed random variables $S_1,S_2$ as
\begin{equation}\label{ncv_a}
    \text{NCV}(S_1,S_2) = \frac{\mathbb{E}(S_1 \text{sgn}(S_2))}{\mathbb{E}|S_2|}.
\end{equation}
Although the analyzed methodology is based on the ACVF measure, we can just replace it with the normalized autocovariation, as presented e.g. in \cite{electronics10151863,kruczek2021generalized}. Its estimator (called sample NCV) for vectors $\mathbf{w}_1,\mathbf{w}_2$ has the following form
\begin{equation}
\lambda(\mathbf{w}_1,\mathbf{w}_2) = \frac{\sum_{i=1}^N w_1^i \text{sgn}(w_2^{i})}{\sum_{i=1}^N |w_2^{i}|}.
\end{equation}
This statistic is also well-defined for complex vectors for which we use $\text{sgn}(z) = z/|z|$ (as in the Kendall method). Let us note that the approach based on NCV is significantly different from the methodology proposed in this paper. Here, we apply different robust estimators of the measures defined in Eqs. (\ref{measure0}) or (\ref{measure}), while the sample NCV estimates the measure given in Eq. (\ref{ncv_a}) which have different properties than the classical measure. 

In the presented simulation study and real data analysis, for the $M_2^c$ (trimmed) estimator, we consider two values of the trimming constant: $c=0.015$ and $c=0.025$. In all spectral coherence maps (for both simulated and real signals), we set nfft=512, Hann window $w(\cdot)$ of length $n=128$ and overlap size nover=110, and $\epsilon_{\min}=3$, $\epsilon_{\max}=100$. Moreover, in order to compare all considered maps for the same scale $[0,1]$, we rescaled each map (we divided each value  by the maximum value of given map). This step is done for both simulated and real signals.

\begin{figure}
    \centering
    \includegraphics[width=\textwidth]{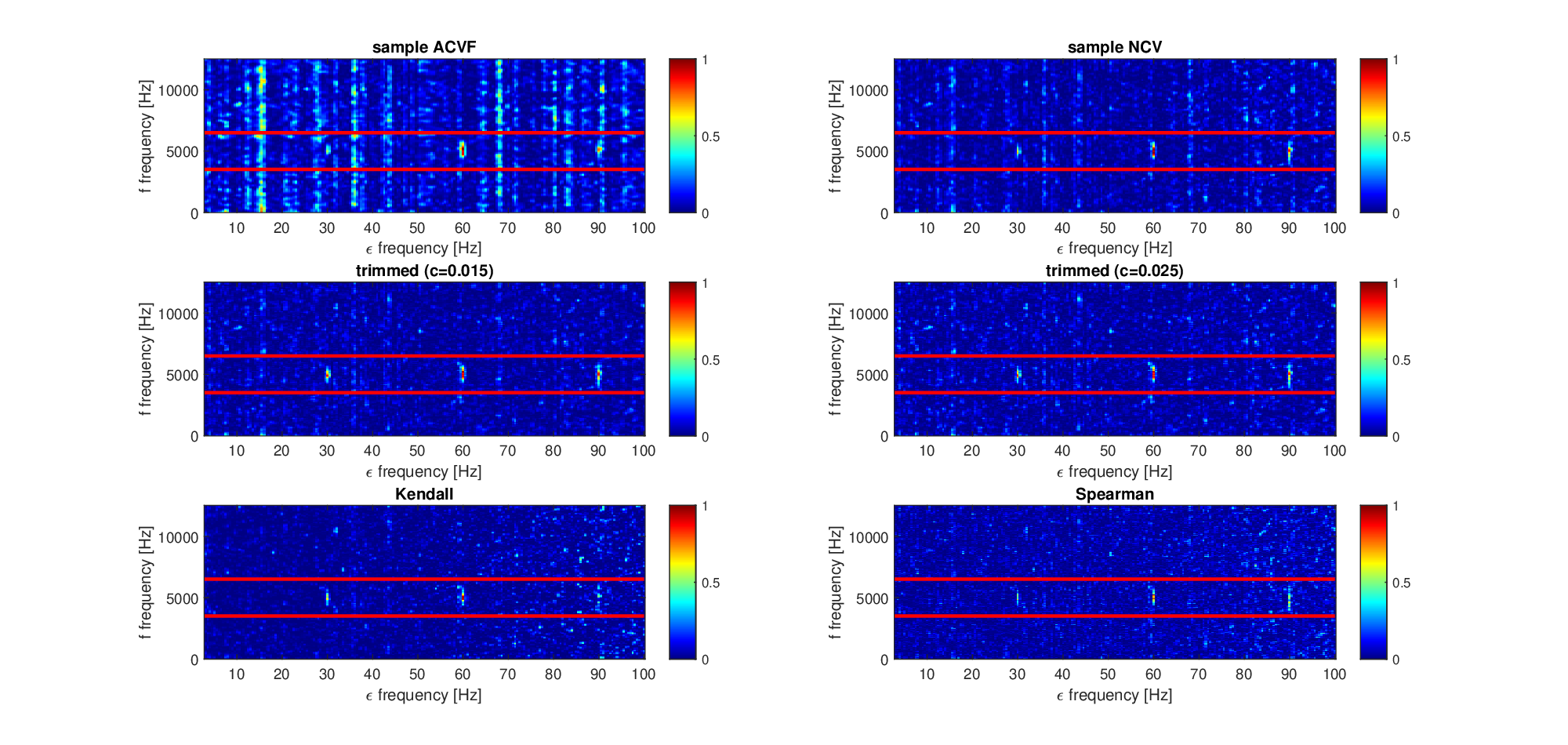}
    \caption{Spectral coherence maps $|\gamma(f,\epsilon)|^2$ for Case 1 ($\{Z_t\} \sim \mathcal{M}(300,0.001,8)$). The true informative frequency band is marked with red lines.}
    \label{ff_case1_map}
\end{figure}

The SC maps calculated for Case 1 are presented in Fig. \ref{ff_case1_map}. For all of them, we expect that the values obtained for both $f \in f_c$ and $\epsilon = 30,60,90$ (further referred to as cyclic $\epsilon$) will be visibly larger than values obtained anywhere else. One can see that in the classical SC map (i.e. based on the sample ACVF) it is not a case. This map is strongly disturbed by non-cyclic impulses present in the signal, which results in significant values obtained for not expected $\epsilon$ frequencies, for the whole range of $f$ frequencies. In the maps for other considered measures, the cyclic impulses are much more visible.

\begin{figure}
    \centering
    \includegraphics[width=\textwidth]{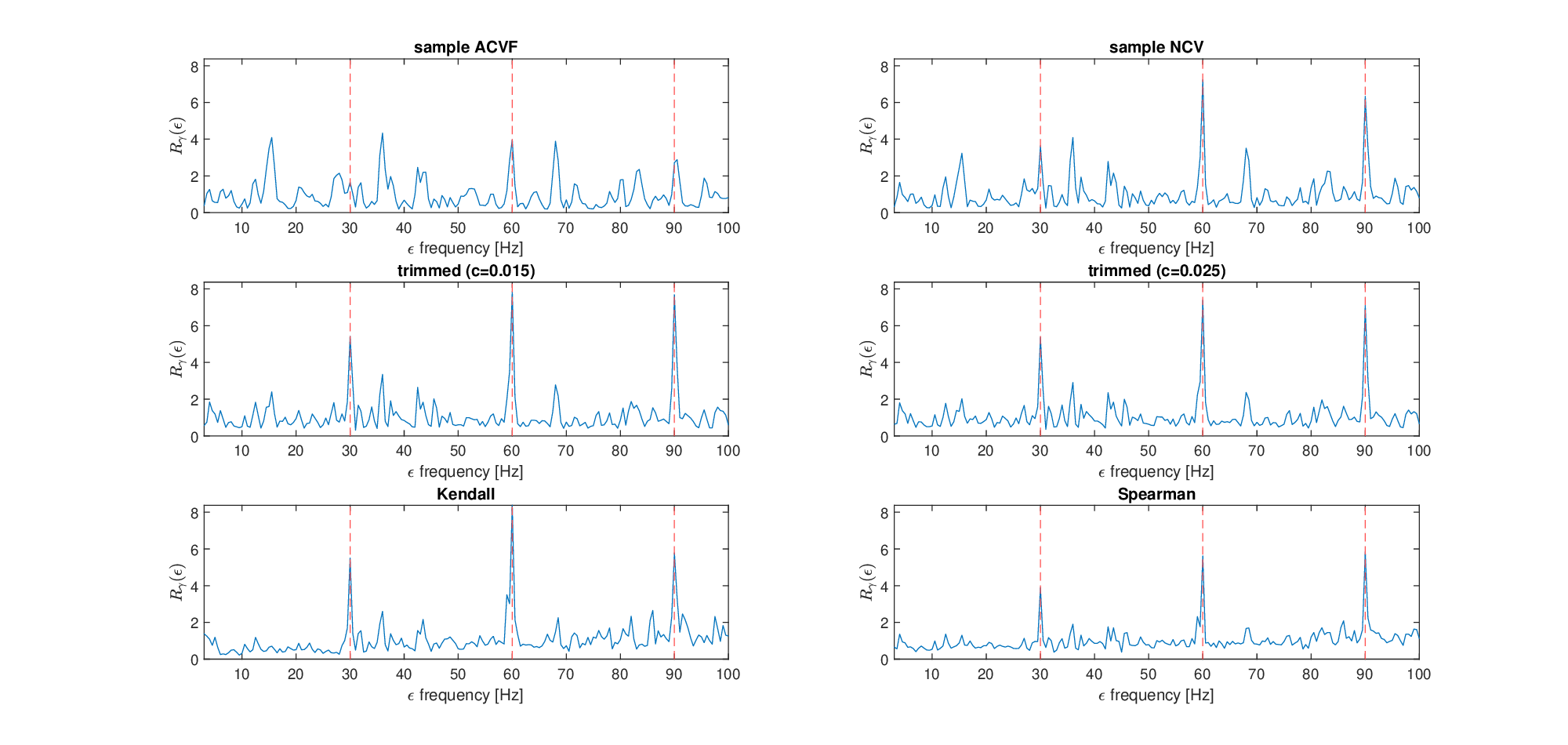}
    \caption{Amplitude ratios $R_\gamma(\epsilon)$ for Case 1 ($\{Z_t\} \sim \mathcal{M}(300,0.001,8)$). The cyclic $\epsilon=30,60,90$ are marked with red dashed lines.}
    \label{ff_case1_ampratio_fixed}
\end{figure}

To assess the quality of the SC maps (i.e. to compare values of periodic impulses in $f_c$ band with the noise on the map), let us consider the amplitude ratio defined as the following function of $\epsilon$
\begin{equation}\label{amp_ratio_gamma}
        R_\gamma(\epsilon) = \frac{\text{mean}\left\{\left|\gamma(f,\epsilon) \right|^2\: : \: f\in f_c\right\}}{\overline{\left|\gamma\right|^2}},
\end{equation}
where $\overline{\left|\gamma\right|^2}$ is the mean of all values of the map. 
The amplitude ratios obtained for Case 1 are illustrated in Fig. \ref{ff_case1_ampratio_fixed}. Of course, their values should be large for $\epsilon=30,60,90$, and close to zero for others. One can see that, for cyclic $\epsilon$, the values for the classical SC map are much lower than for sample NCV and the proposed robust approaches. Moreover, they are not larger than values for other $\epsilon$ for this map, so that one cannot detect the actual periodicity. On the other hand, in the amplitude ratios for other approaches, one can clearly identify the cyclic behavior. 

\begin{figure}
    \centering
    \includegraphics[width=\textwidth]{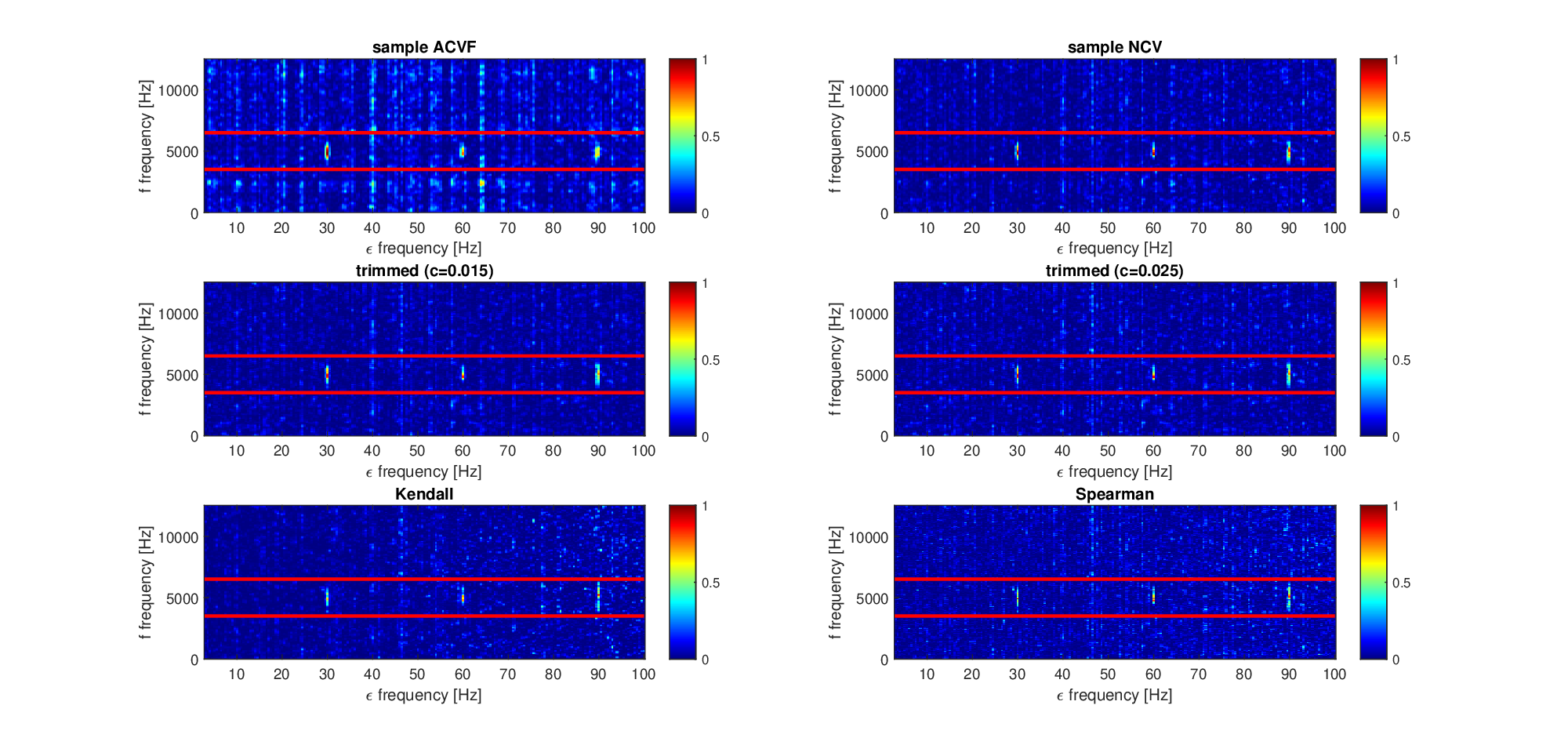}
    \caption{Spectral coherence maps $|\gamma(f,\epsilon)|^2$ for Case 2 ($\{Z_t\} \sim \mathcal{T}(2,3)$). The true informative frequency band is marked with red lines.}
    \label{ff_case2_map}
\end{figure}

\begin{figure}
    \centering
    \includegraphics[width=\textwidth]{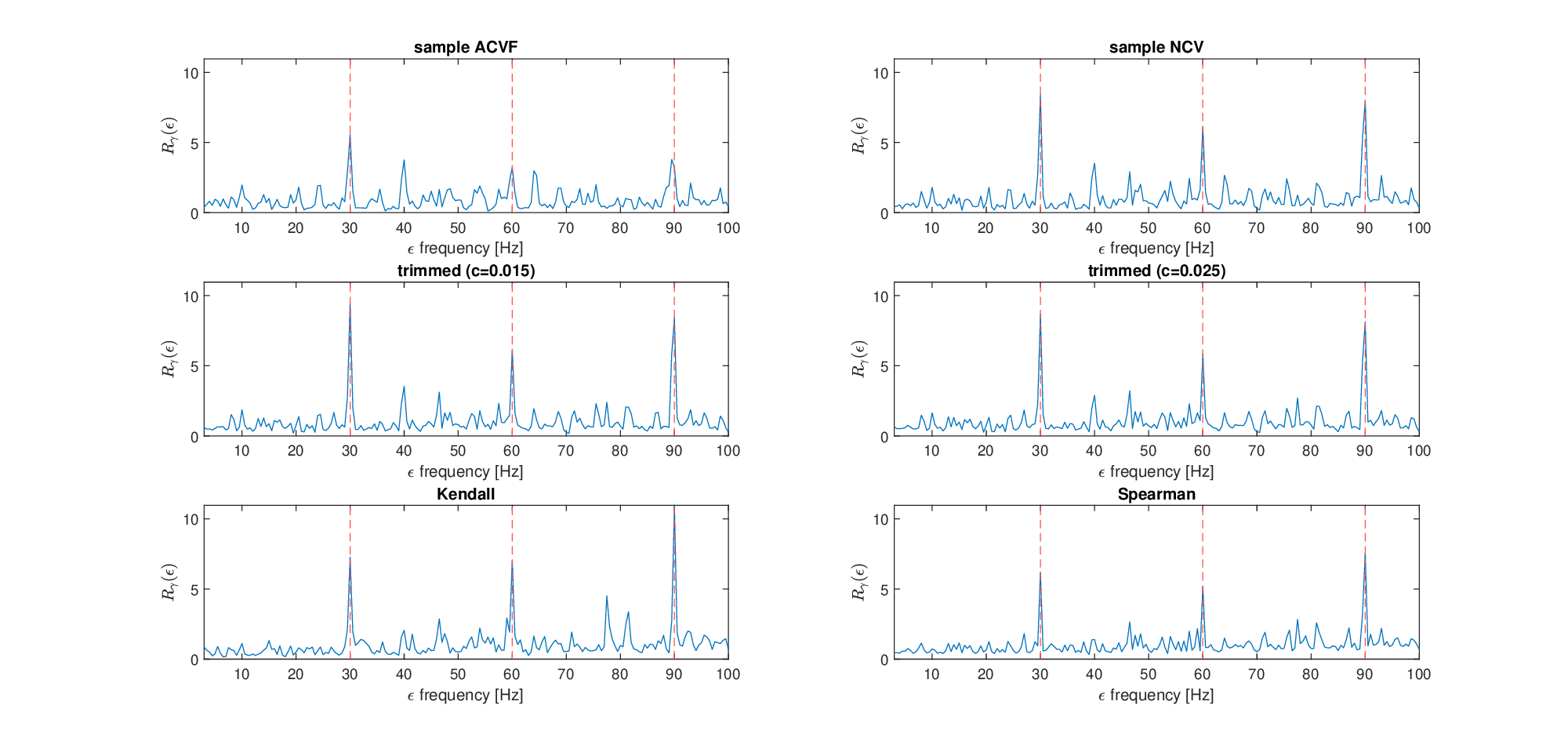}
    \caption{Amplitude ratios $R_\gamma(\epsilon)$ for Case 2 ($\{Z_t\} \sim \mathcal{T}(2,3)$). The cyclic $\epsilon=30,60,90$ are marked with red dashed lines.}
    \label{ff_case2_ampratio_fixed}
\end{figure}

The spectral coherence maps for Case 2 are presented in Fig. \ref{ff_case2_map}. In all maps, the cyclic impulses are visible. However, in the sample ACVF-based map, it is more difficult to identify them because of the relatively high values in the background noise. The corresponding amplitude ratios are illustrated in Fig. \ref{ff_case2_ampratio_fixed}. Again, the robust and NCV-based approaches indicate the periodicity much better than the classical map.  

\begin{figure}
    \centering
    \includegraphics[width=\textwidth]{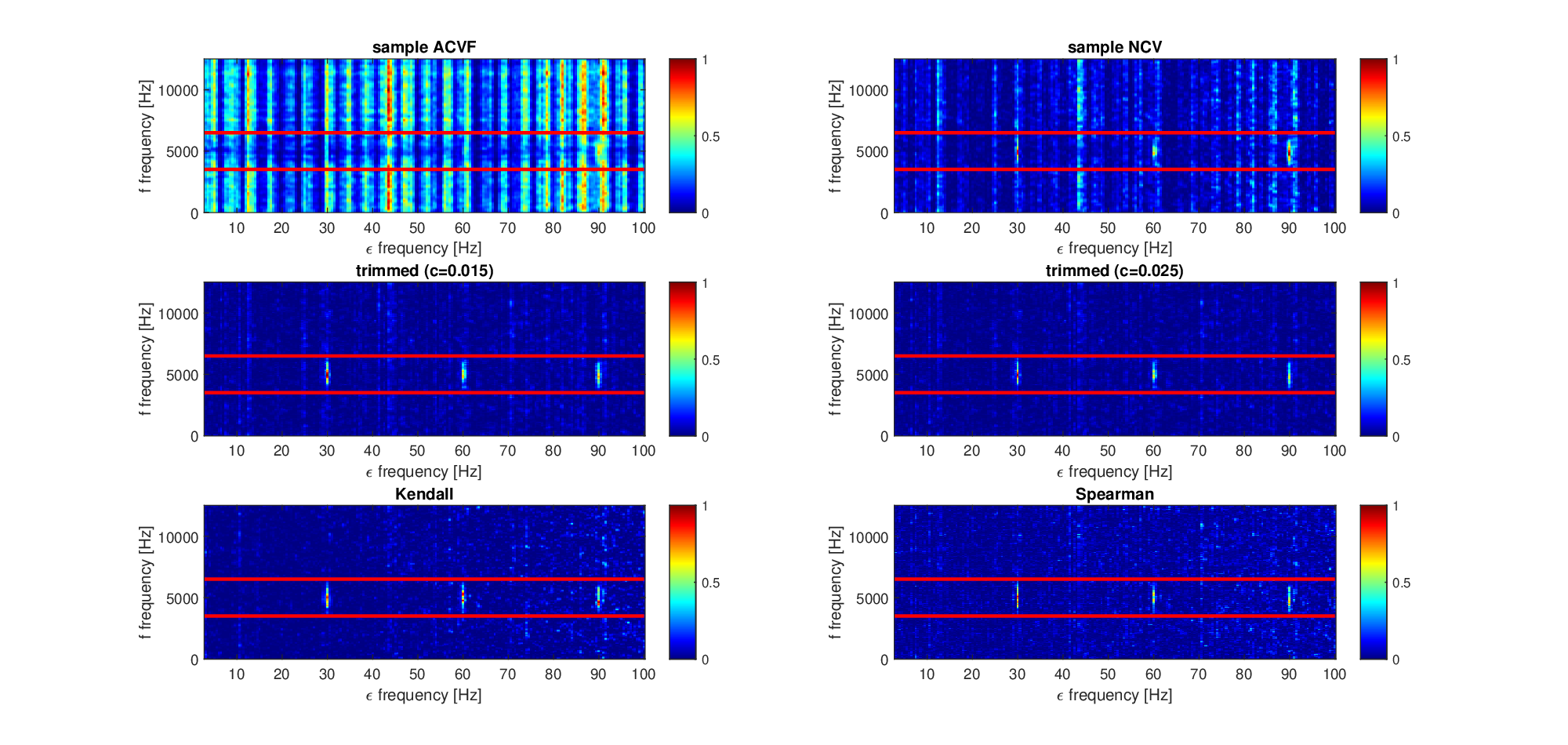}
    \caption{Spectral coherence maps $|\gamma(f,\epsilon)|^2$ for Case 3 ($\{Z_t\} \sim \mathcal{S}(1.7,3)$). The true informative frequency band is marked with red lines.}
    \label{ff_case3_map}
\end{figure}

\begin{figure}
    \centering
    \includegraphics[width=\textwidth]{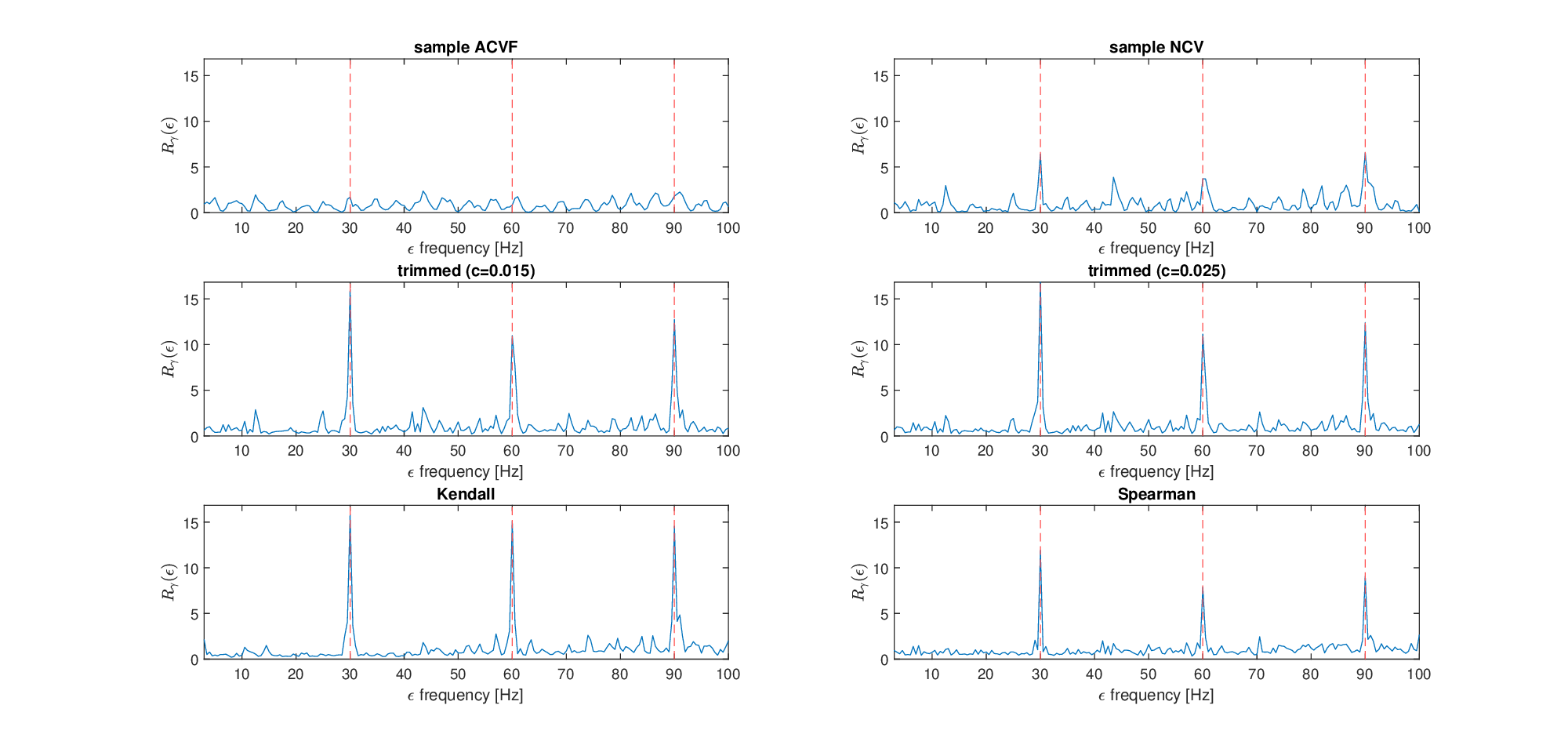}
    \caption{Amplitude ratios $R_\gamma(\epsilon)$ for Case 3 ($\{Z_t\} \sim \mathcal{S}(1.7,3)$). The cyclic $\epsilon=30,60,90$ are marked with red dashed lines.}
    \label{ff_case3_ampratio_fixed}
\end{figure}

In Case 3, the advantage of the proposed robust approaches is the largest. As one can see on the maps constructed for this case and presented in Fig. \ref{ff_case3_map}, the classical sample ACVF-based map is strongly contaminated, and the expected periodicity is not identifiable at all. Moreover, although in the NCV-based map the cyclic impulses are present, they are barely visible because of the large values in the background noise. On the other hand, on robust maps, the observed behavior is much closer to the expected one. The values in informative frequency band and cyclic frequencies are significantly larger than the other ones on the map -- hence, the periodicity is easily detectable. This is confirmed by amplitude ratios calculated for this case, presented in Fig. \ref{ff_case3_ampratio_fixed}. Let us note that the results obtained for the classical SC map do not indicate any periodicity.

\begin{figure}
    \centering
    \includegraphics[width=0.4\textwidth]{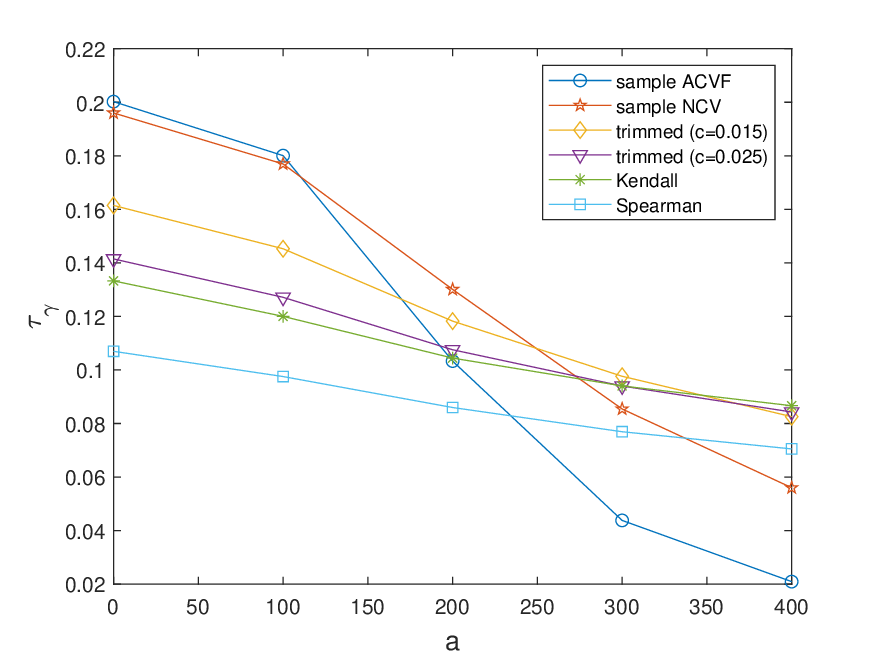}
     \includegraphics[width=0.4\textwidth]{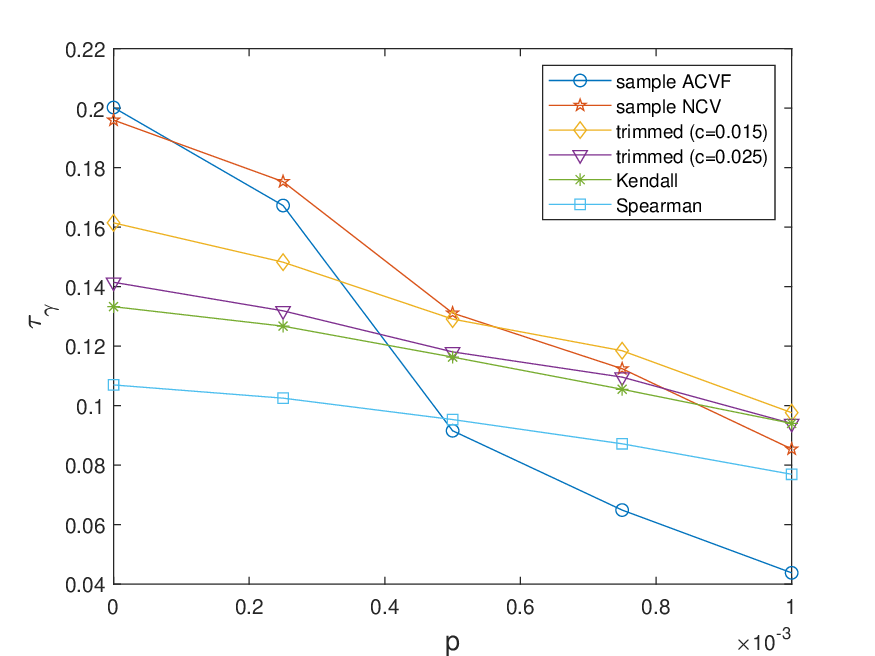}
    \caption{Performance indicator $\tau_\gamma$ values for different cases of $a$ (left panel) with $\{Z_t\} \sim \mathcal{M}(a,0.001,8)$ and $p$ (right panel) with $\{Z_t\} \sim \mathcal{M}(300,p,8)$.}
    \label{ff_different_M}
\end{figure}

To analyze the performance for different parameters of the $\{Z_t\}$ distribution, we define the performance indicator as follows
\begin{equation}\label{ind_gamma}
        \tau_\gamma = \frac{\sum_{\epsilon \text{ cyclic}} R_\gamma(\epsilon)}{\sum_{\epsilon=\epsilon_{min}}^{\epsilon_{max}} R_\gamma(\epsilon)}.
\end{equation}
In the left panel of Fig. \ref{ff_different_M}, we present the $\tau_\gamma$ obtained for different values of $a$ parameter, assuming $\{Z_t\} \sim \mathcal{M}(a,0.001,8)$. In the right panel of Fig. \ref{ff_different_M}, analogously, we present the indicator values for different $p$ with $\{Z_t\} \sim \mathcal{M}(300,p,8)$. One can see that in extreme cases (i.e. for larger $a$ or $p$) the robust approaches outperform the sample ACVF- and NCV-based methods.  

\begin{figure}
    \centering
    \includegraphics[width=0.4\textwidth]{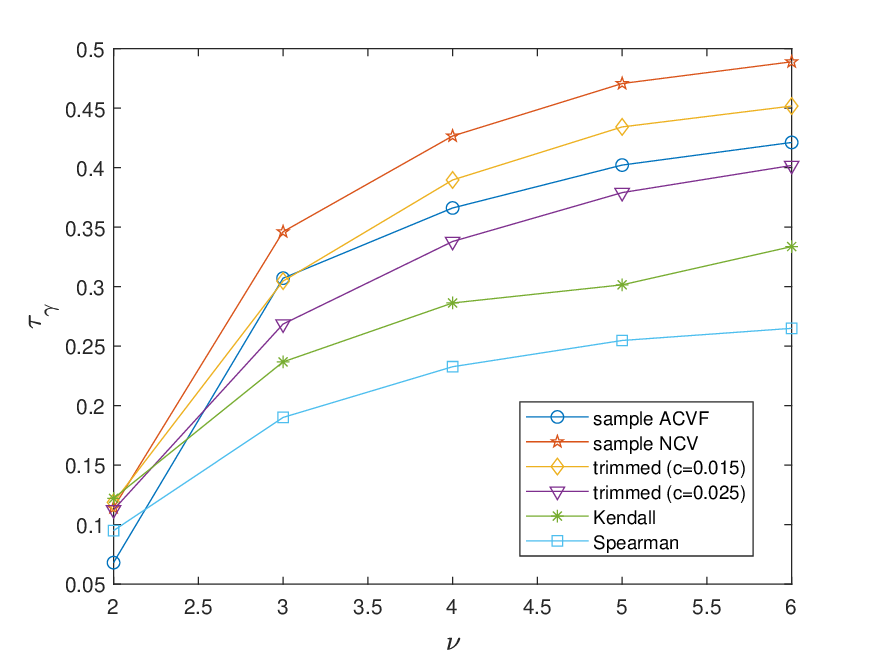}
     \includegraphics[width=0.4\textwidth]{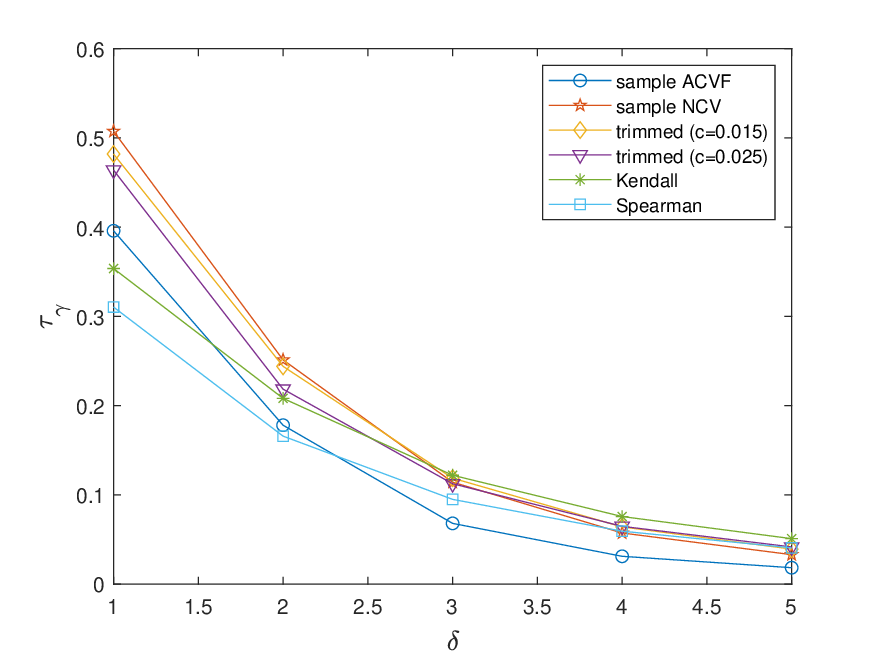}
    \caption{Performance indicator $\tau_\gamma$ values for different cases of $\nu$ (left panel) with $\{Z_t\} \sim \mathcal{T}(\nu,3)$ and $\delta$ (right panel) with $\{Z_t\} \sim \mathcal{T}(2,\delta)$.}
    \label{ff_different_T}
\end{figure}

In a similar way we consider other analyzed distributions. In Fig. \ref{ff_different_T}, we present the indicator values for different $\nu$ (left panel) with $\{Z_t\}\sim\mathcal{T}(\nu,3)$, and for different $\delta$ (right panel) with $\{Z_t\} \sim \mathcal{T}(2,\delta)$. Let us note that in this case, perhaps surprisingly, the best results in general were obtained for sample NCV-based method. However, again for most extreme set-ups (for $\nu=2$ and larger $\delta$), proposed robust approaches have a slight advantage.

\begin{figure}
    \centering
    \includegraphics[width=0.4\textwidth]{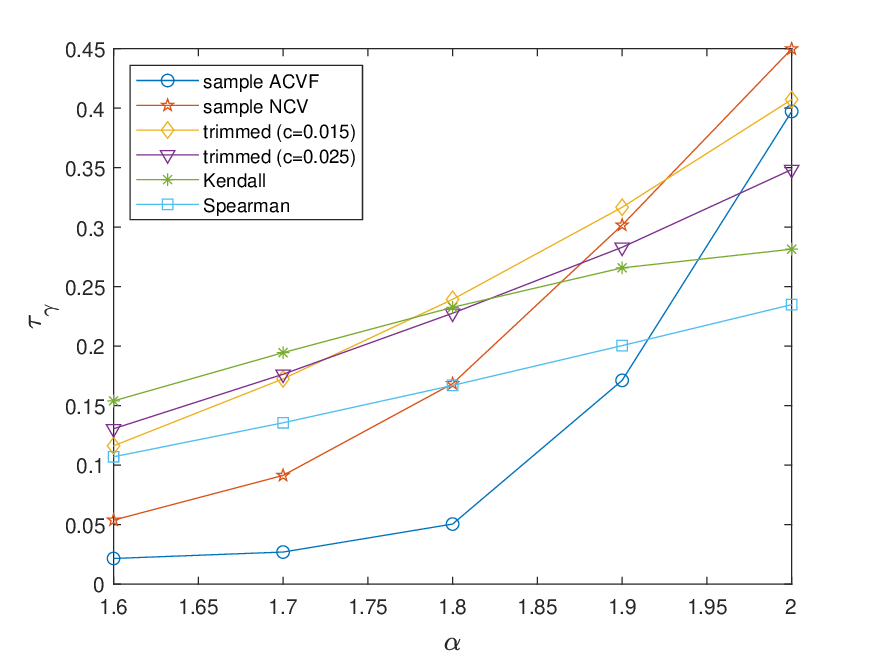}
     \includegraphics[width=0.4\textwidth]{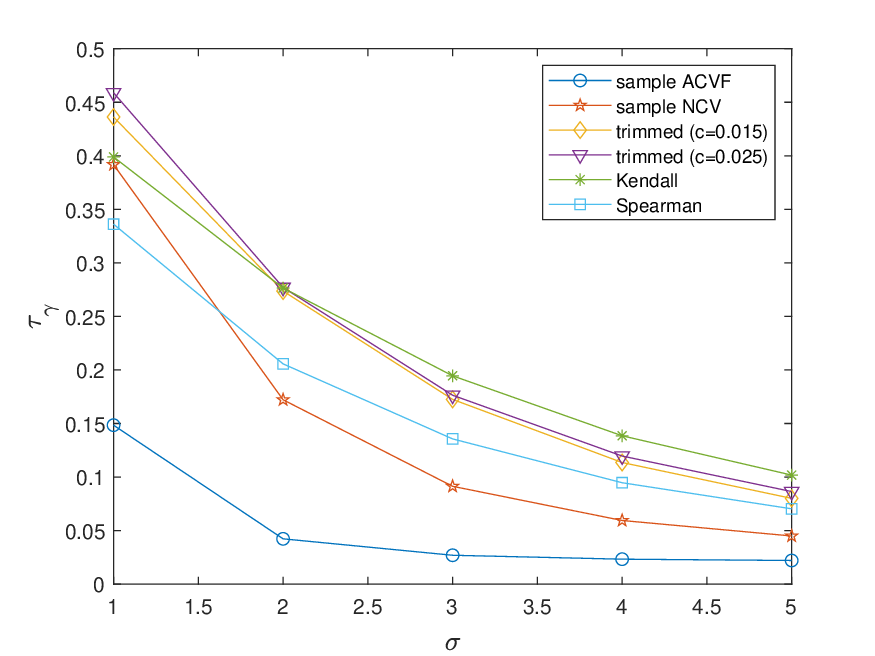}
    \caption{Performance indicator $\tau_\gamma$ values for different cases of $\alpha$ (left panel) with $\{Z_t\} \sim \mathcal{S}(\alpha,3)$ and $\sigma$ (right panel) with $\{Z_t\} \sim \mathcal{S}(1.7,\sigma)$.}
    \label{ff_different_S}
\end{figure}

In Fig. \ref{ff_different_S}, we present the results for different $\alpha$ (left panel) with $\{Z_t\}\sim\mathcal{S}(\alpha,3)$, and for different $\sigma$ (right panel) with $\{Z_t\} \sim \mathcal{S}(1.7,\sigma)$. Here, for small values of $\alpha$, one can clearly see an advantage of the proposed robust methods over the other ones. The same holds for all considered $\sigma$, as here we also have a fairly low $\alpha=1.7$. Let us also note that the decline of performance (for lower $\alpha$ and larger $\sigma$) for the Kendall correlation-based map is clearly slower than for other methods. In fact, it turns out to be the best one for extreme cases. Similar behavior can be observed also in previously considered cases of $\{Z_t\}$ distribution. 

Although in this paper we focus on the problem of periodicity detection and the performance of the considered methods in this particular task, the analysis of the behavior of the maps for a signal without periodicity is also important. Let us note that such a signal would indicate the healthy state of a machine.  In \ref{appendix_clean}, we present and shortly discuss the results for the signals from Cases 1-3 without cyclic impulses (i.e., for $X_t = Z_t$ in Eq. \eqref{model2}). The presented results clearly confirm that the robust maps (in particular for Kendall and Spearman estimators) are much less disturbed than the classical one, which indicates the superiority of the proposed approach over the classical one. 

\section{Real signals analysis}\label{real}

\begin{figure}
    \centering
    \includegraphics[width=0.4\textwidth]{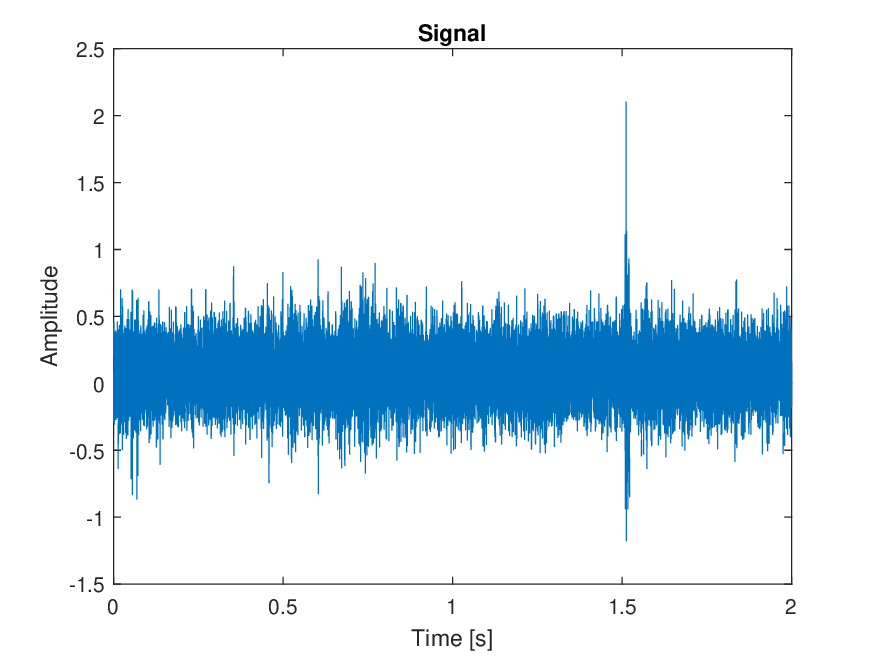}
     \includegraphics[width=0.4\textwidth]{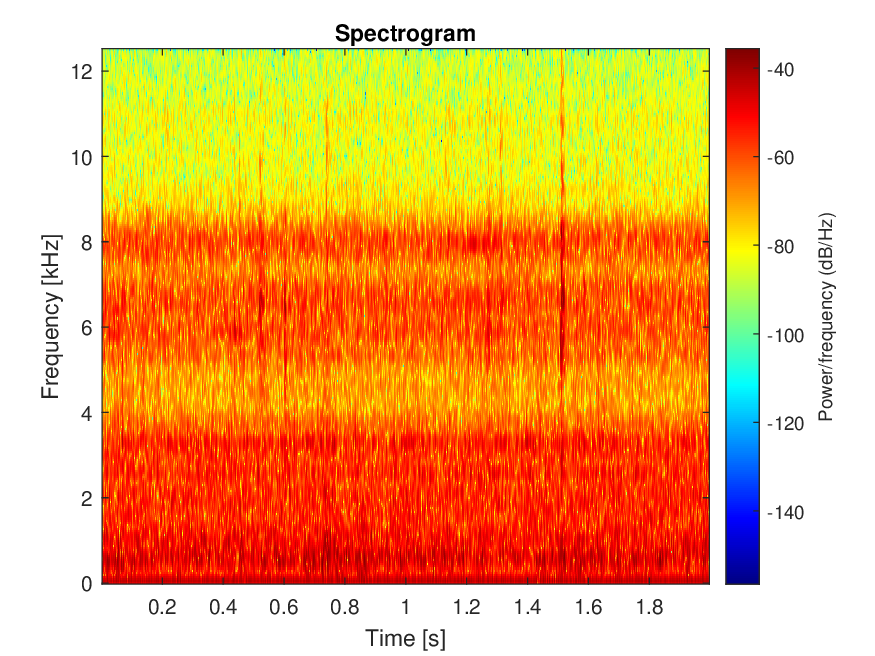}
    \caption{The analyzed real vibration signal from a crushing machine (Dataset 1) and its spectrogram.}
    \label{signal_real}
\end{figure}

In this section, we present the application of the considered methods to a real vibration dataset from a crushing machine. It is the two-second sample selected from the signal analyzed in \cite{kruczek2021generalized}. The system used for measurement consists of Endevco accelerometers (vibration), BruelKjaer Laser probe (shaft speed was measured), NI DAQ card, Labview Signal Express and notebook. The vibration signals were acquired in horizontal and vertical directions. Similarly as in the signals presented in Section \ref{sec:sim}, we have $f_s = 25000$ Hz and hence $L=50000$. The analyzed data were collected from a healthy machine; thus, their analysis should not indicate any periodicity. The signal and its spectrogram are presented in Fig. \ref{signal_real}. Most of all, let us note the presence of several significantly outlying values, in particular around 1.5 s timestamp. Obviously, this observation motivates us to apply the proposed robust methods. 

\begin{figure}
    \centering
    \includegraphics[width=0.4\textwidth]{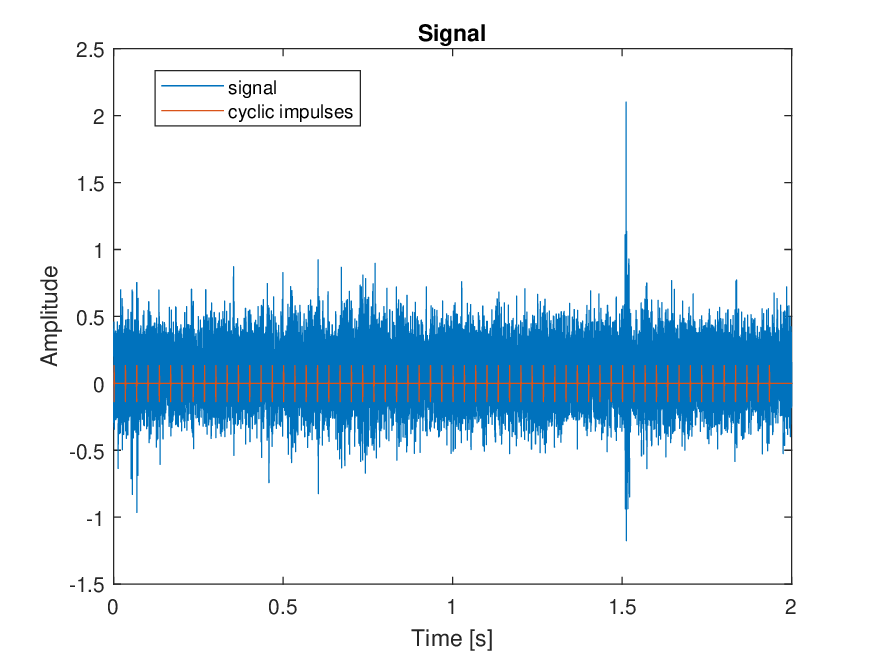}
     \includegraphics[width=0.4\textwidth]{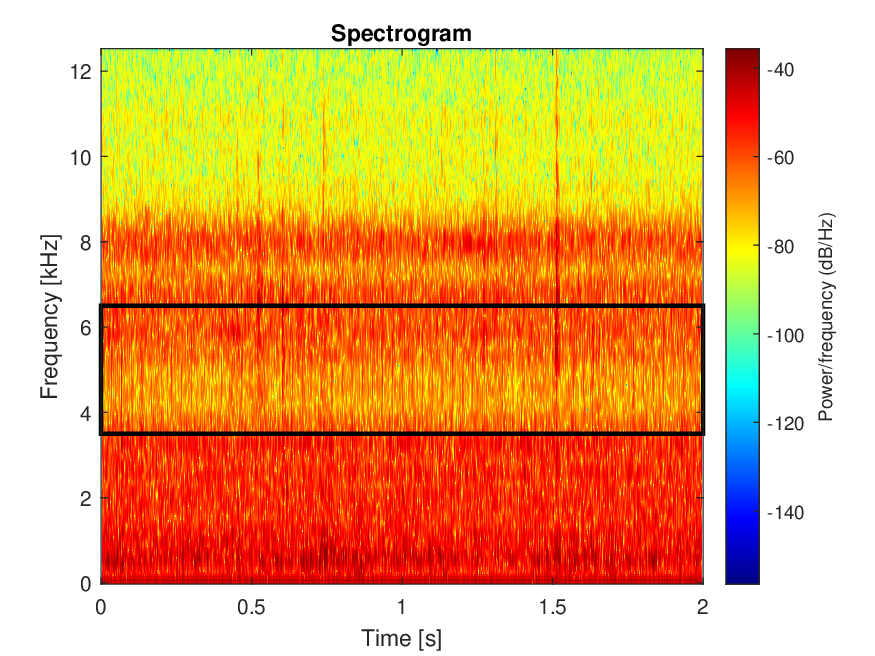}
    \caption{The modified vibration signal with added periodic impulses (Dataset 2) and its spectrogram.}
    \label{signal_real_modified}
\end{figure}

For the analysis of the presented data the analyzed bi-frequency methodology is utilized. The conducted procedure is as follows. First, we apply all six considered variants of the given methodology (the same as in Section \ref{sec:sim}, i.e. using sample ACVF, sample NCV, trimmed with $c=0.015$ and $c=0.025$, Kendall and Spearman estimators) to the data presented in Fig. \ref{signal_real}, further referred to as Dataset 1. We present the results in \ref{appendix_clean}, in a similar way as for the non-periodic versions of Cases 1-3. Then, we consider a modified signal, adding artificial periodic impulses to the original dataset. To such data (denoted later as Dataset 2), we again apply the considered methods, in order to assess their efficiency in the periodicity identification. The periodic impulses have the same structure (see Eq. \eqref{imp_structure}) as these included in the signals analyzed in Section \ref{sec:sim}, only for a different amplitude $B=0.25$. The modified signal and its spectrogram are presented in Fig. \ref{signal_real_modified}. One can see that the included periodic impulses are relatively low in amplitude, particularly in comparison with mentioned outliers. At the end, we present the performance indicator $\tau_\gamma$ results for different amplitudes $B = 0.125,0.25,0.375,0.5$ of the periodic impulses added to the original signal. 

The results for the Dataset 1 (see \ref{appendix_clean}, Fig. \ref{ff_dataset1_map} and Fig. \ref{ff_dataset1_amp}) show that robust approaches (in particular, Kendall and Spearman methods) produced significantly smaller disturbances in the maps than the classical one.  Most of all, in the sample ACVF-based map, the results for $f>5000$ Hz are significantly larger than for lower $f$ frequencies, which is an undesired behavior for this signal.

\begin{figure}
    \centering
    \includegraphics[width=\textwidth]{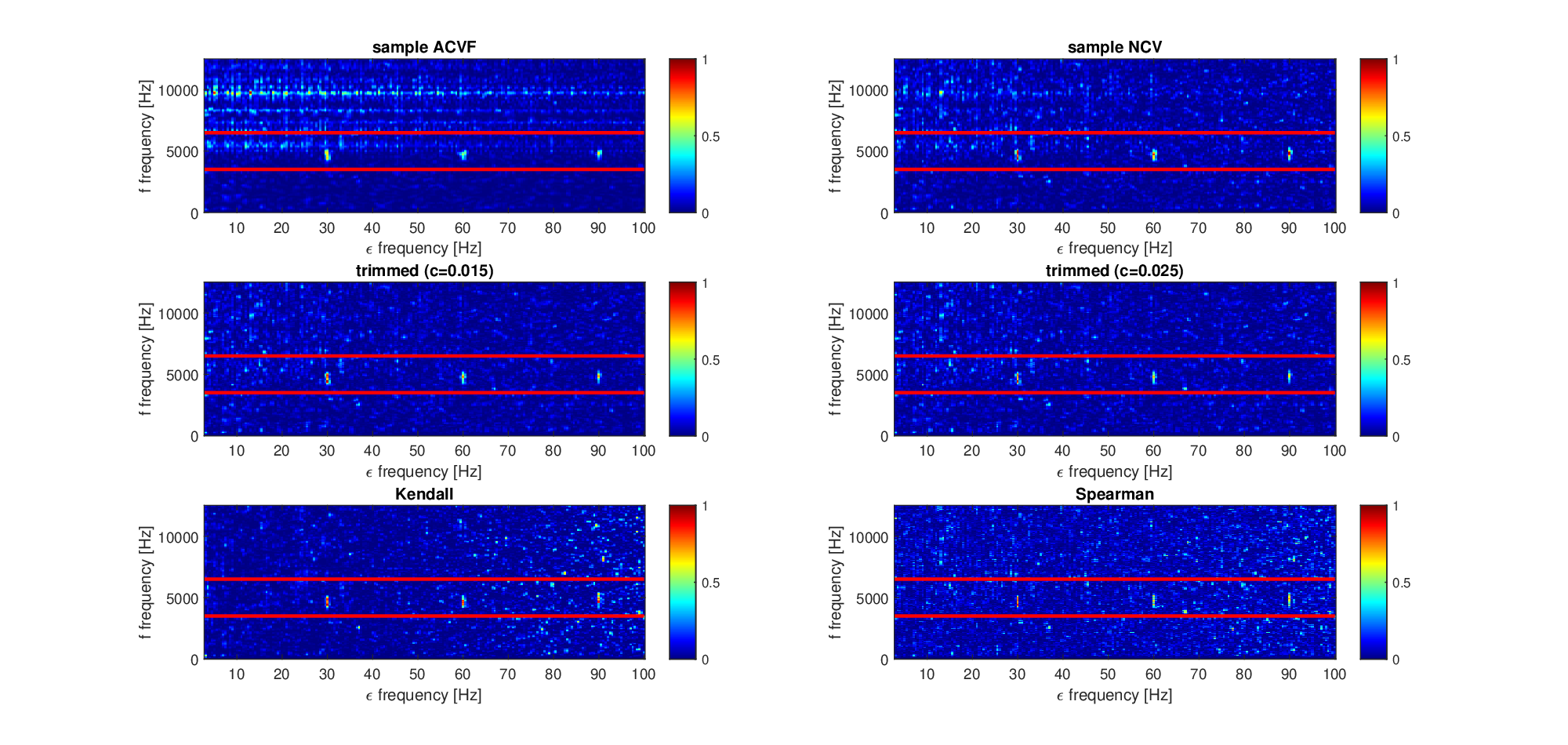}
    \caption{Spectral coherence maps $|\gamma(f,\epsilon)|^2$ for the Dataset 2 (real vibration signal with artificial periodic impulses). The true informative frequency band is marked with red lines.}
    \label{ff_dataset2_map}
\end{figure}

\begin{figure}
    \centering
    \includegraphics[width=\textwidth]{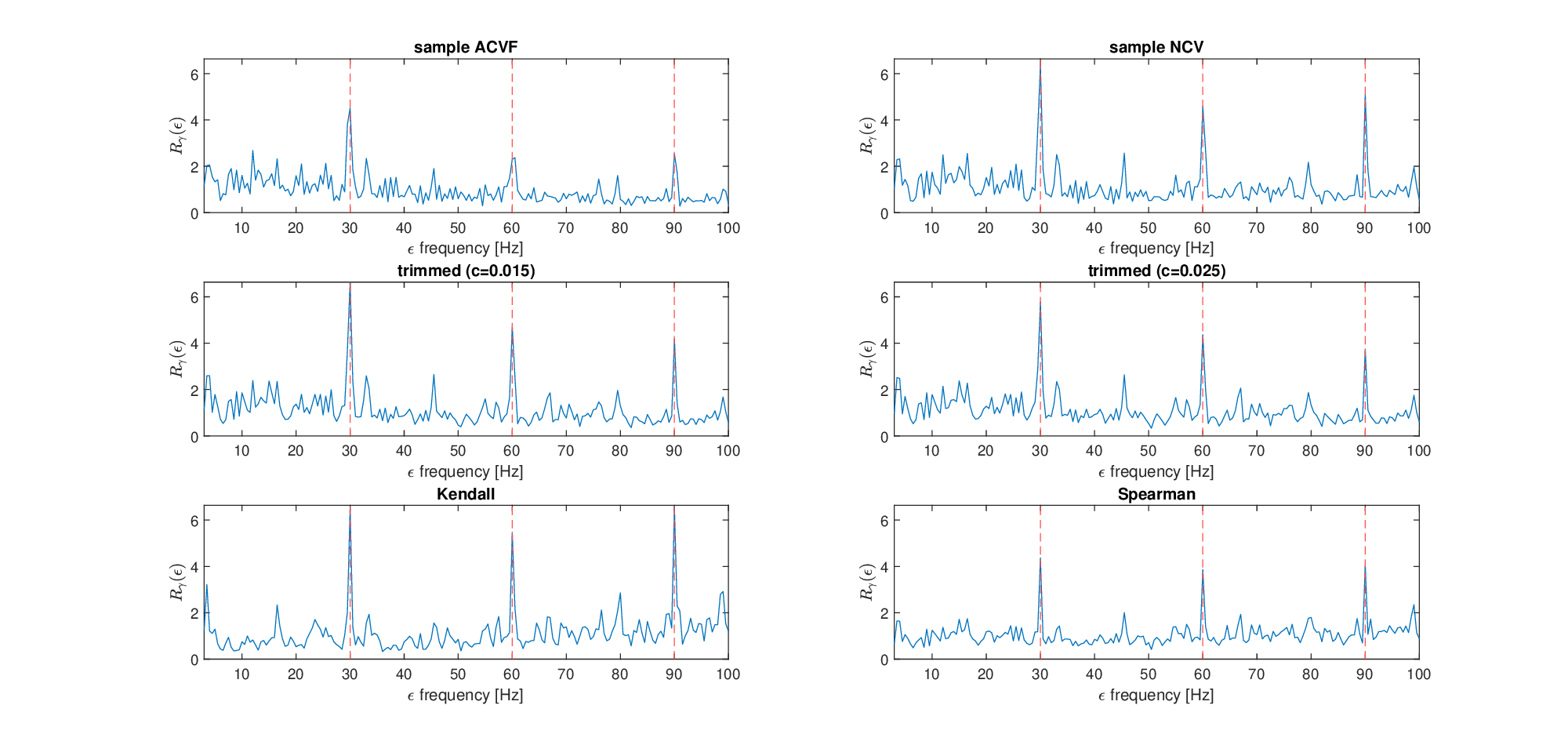}
    \caption{Amplitude ratios $R_\gamma(\epsilon)$ for the Dataset 2 (real vibration signal with artificial periodic impulses). The cyclic $\epsilon=30,60,90$ are marked with red dashed lines.}
    \label{ff_dataset2_ampratio_fixed}
\end{figure}

\begin{figure}
    \centering
    \includegraphics[width=0.5\textwidth]{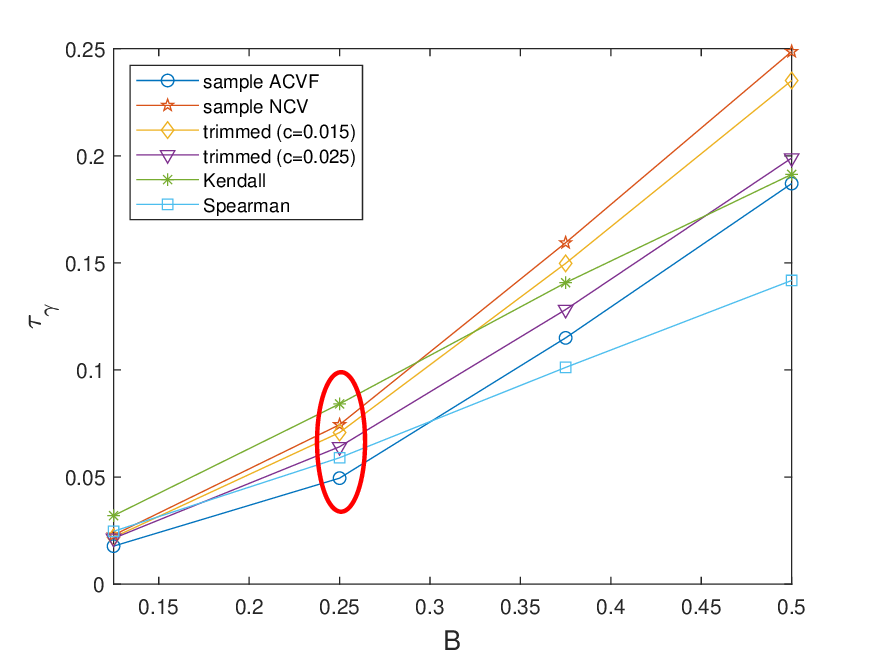}
    \caption{Performance indicator $\tau_\gamma$ values for different amplitudes $B$ of the periodic impulses added to the real vibration signal (the case denoted as Dataset 2 is marked as red ellipse).}
    \label{ff_different_B}
\end{figure}

The spectral coherence maps for the periodic signal Dataset 2 are presented in Fig. \ref{ff_dataset2_map}. All considered methods were able to detect the periodicity. However, in the sample ACVF-based map it is slightly less clearly visible because of the behavior of its background noise (that is, large values for higher $f$). In the robust maps, as well as in the NCV-based one, the expected periodic impulses are better identifiable. The amplitude ratios in Fig. \ref{ff_dataset2_ampratio_fixed} and performance indicators $\tau_\gamma$ in Tab. \ref{tab:ff_dataset2_taugamma} confirm this observation. In Fig. \ref{ff_different_B}, the performance indicators for different amplitudes of added periodic impulses are presented. For all methods, the larger the amplitude, the better the efficiency. One can see that for $B=0.125$ and $B=0.25$ the best result was obtained using the Kendall estimator. On the other hand, for $B=0.375$ and $B=0.5$, the NCV and trimmed estimator for $c=0.015$ were the most efficient. Let us note that in all considered cases, the standard sample ACVF-based method was either the worst or the second worst (only ahead of Spearman estimator).

\begin{table}
    \centering
    \begin{tabular}{|c|c|c|c|c|c|c|} \hline
      method &  sample ACVF & sample NCV & trimm. ($c=0.015$) & trimm. ($c=0.025$) & Kendall & Spearman \\ \hline
       $\tau_\gamma$ & 0.0495 & 0.0744 & 0.0708 & 0.0641 & 0.0841 & 0.0590\\ \hline
    \end{tabular}
    \caption{Performance indicator $\tau_\gamma$ values obtained for Dataset 2.}
    \label{tab:ff_dataset2_taugamma}
\end{table}

\newpage
\section{Conclusions}
This work has focused on the issue of identifying periodic components in signals with heavily-tailed distributed noise. Although the examined problem is discussed in the context of local damage detection, it may be viewed as a universal issue that could occur in a number of applications. The proposed methodology is based on the spectral coherence-based analysis and utilizes the APC technique for SC map calculation. The idea is based on the replacement of the sample  estimator of the ACVF/ACF in the classical algorithm by its robust versions. The robust estimators of the classical measures are less sensitive to large observations resulting from non-Gaussian distribution of the noise, however they still approximate well the theoretical ACVF/ACF.

The efficiency of the new approach has been demonstrated for simulated signals for three selected heavy-tailed distributions with different non-Gaussianity levels. Moreover, a real signal from a crushing machine has been analyzed. The methods were tested for signals with and without cyclic impulses (corresponding to a damaged and a healthy machine, respectively). The results for both simulated and real data are contrasted with the autocovariation measure specifically designed for $\alpha$-stable distributed signals and the generalized spectral coherence methodology. The presented simulation studies and results for real vibration data have clearly confirmed that the proposed approach outperforms the classical ones. We believe that the introduced algorithm for robust version of CSC can be useful for various real applications.

\section*{Acknowledgements}
The work of WŻ, RZ and AW is supported by National Center of Science under Sheng2 project No. UMO-2021/40/Q/ST8/00024 "NonGauMech - New methods of processing non-stationary signals (identification, segmentation, extraction, modeling) with non-Gaussian characteristics for the purpose of monitoring complex mechanical structures". 
\bibliography{mybibliography}

\appendix
\section{Results for signals without periodicity}\label{appendix_clean}

In this part, we present the results for the signals from Sections \ref{sec:sim} and \ref{real} without added cyclic impulses $s(t)$. 
As currently there is no periodicity in the data, the maps should not indicate it. Moreover, the maps also should not be disturbed by a large number of significant values. 
To analyze the performance of a given map, we calculate the following indicator 
\begin{eqnarray}\label{rprim}
R^\prime_\gamma(\epsilon) = \text{mean}\left\{\left|\gamma(f,\epsilon) \right|^2\right\}.
\end{eqnarray}
{In other words, $R^\prime_\gamma(\epsilon)$ is the mean of map values from the column corresponding to $\epsilon$.}
For the considered signals (i.e. without periodicity), the indicator (\ref{rprim}) should take the lowest possible values for all $\epsilon$.

\begin{figure}
    \centering
    \includegraphics[width=\textwidth]{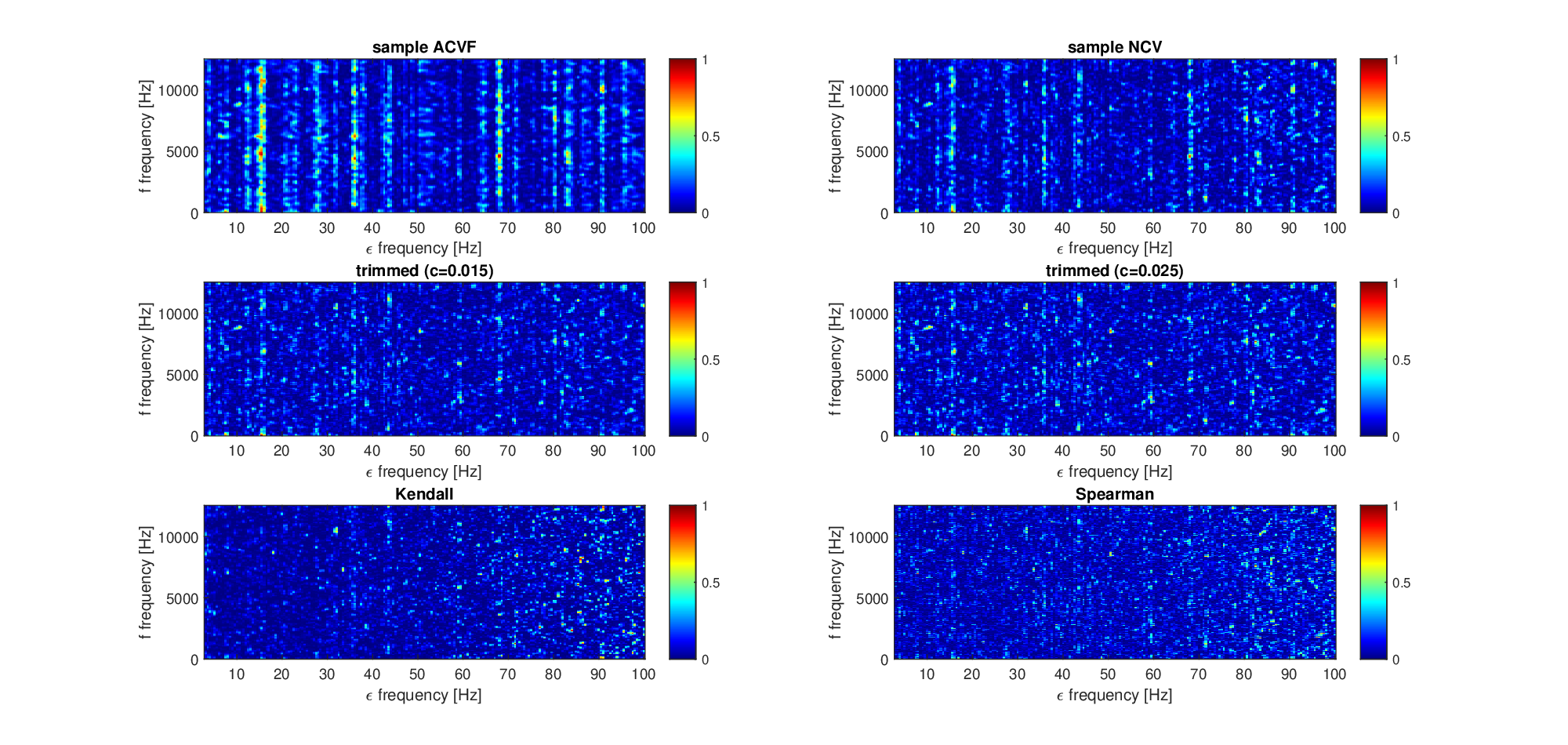}
    \caption{Spectral coherence maps $|\gamma(f,\epsilon)|^2$ for Case 1 without cyclic impulses ($\{Z_t\} \sim \mathcal{M}(300,0.001,8)$).}
    \label{clean_case1_mapa}
\end{figure}

\begin{figure}
    \centering
    \includegraphics[width=\textwidth]{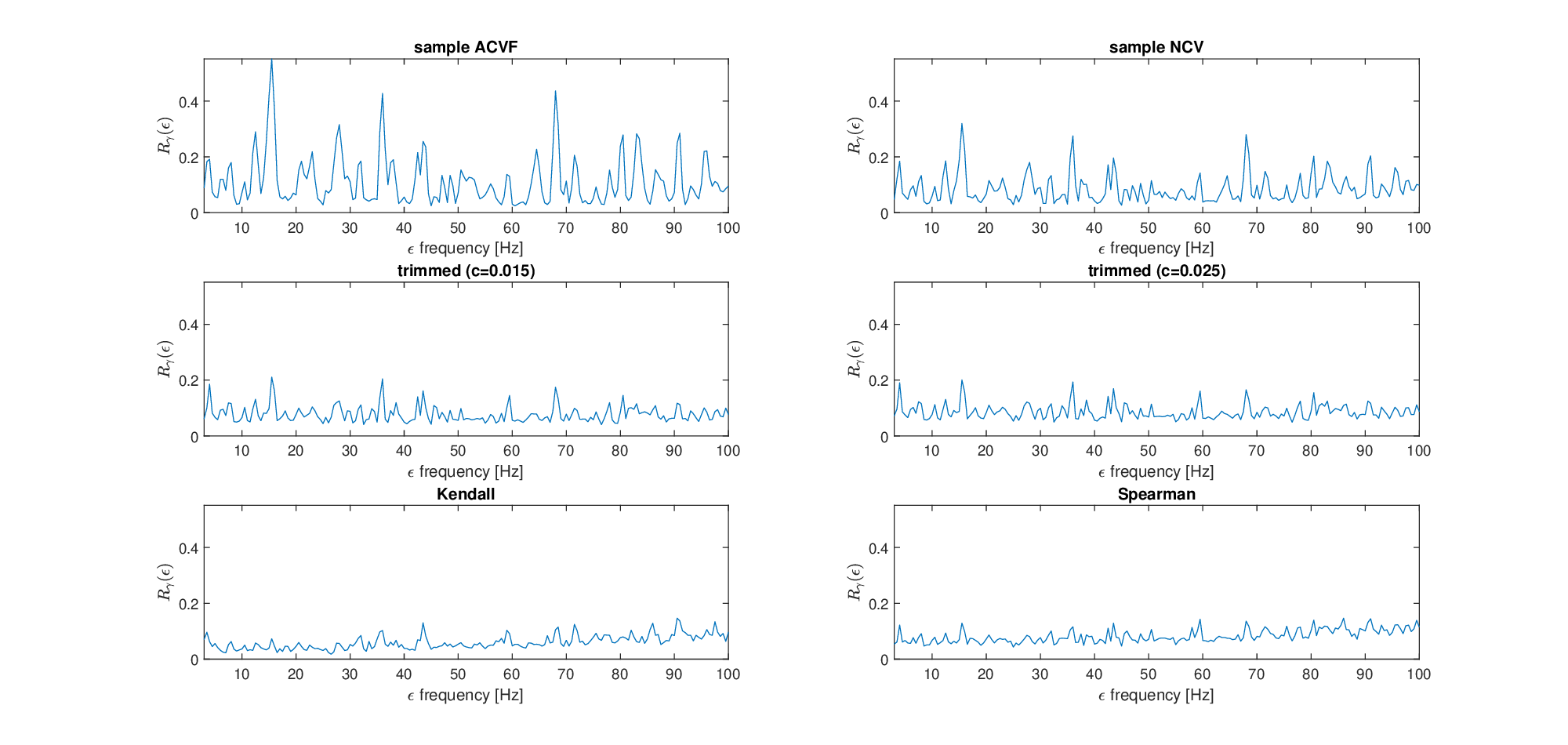}
    \caption{Statistics $R^\prime_\gamma(\epsilon)$ for Case 1 without cyclic impulses ($\{Z_t\} \sim \mathcal{M}(300,0.001,8)$).}
    \label{clean_case1_amp}
\end{figure}

\begin{figure}
    \centering
    \includegraphics[width=\textwidth]{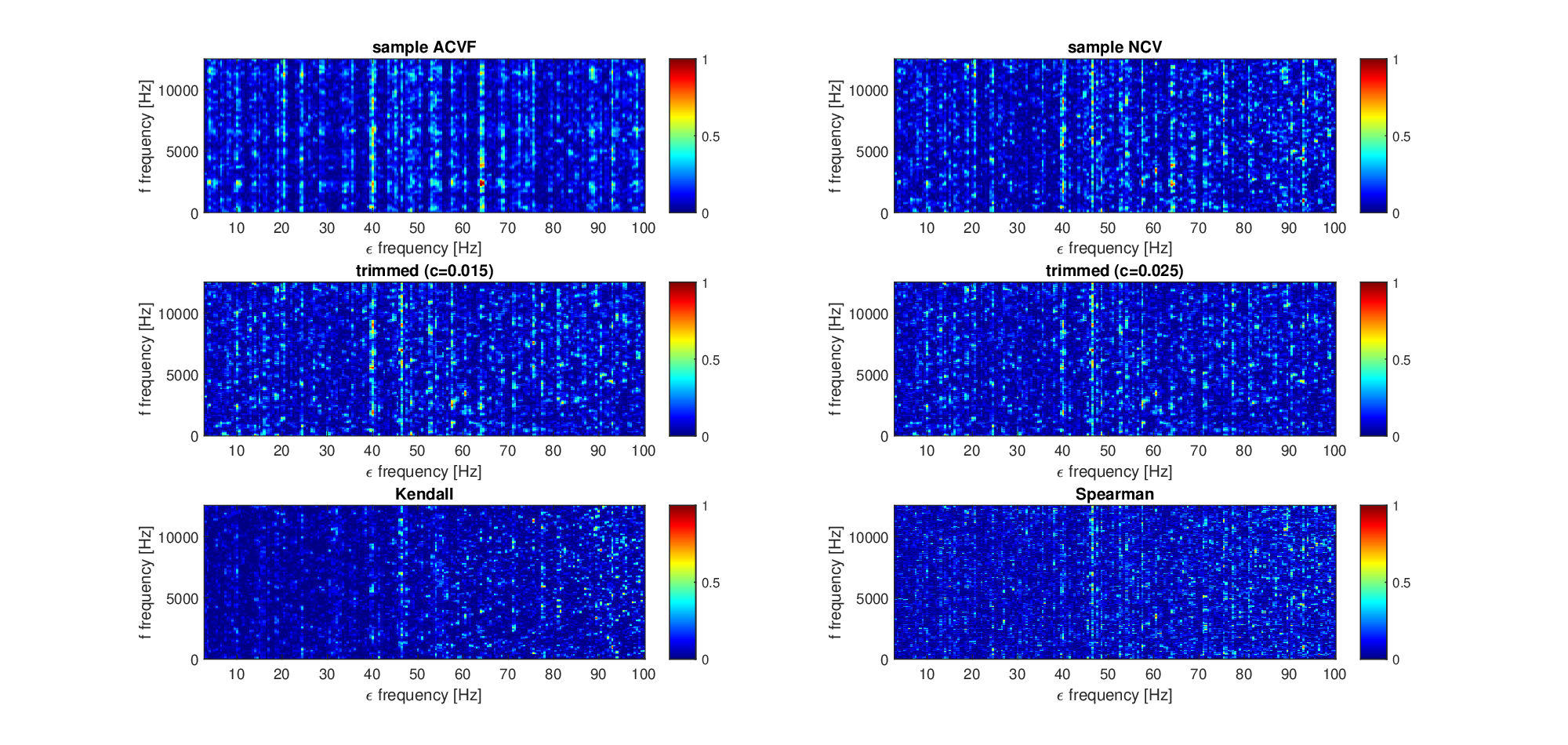}
    \caption{Spectral coherence maps $|\gamma(f,\epsilon)|^2$ for Case 2 without cyclic impulses ($\{Z_t\} \sim \mathcal{T}(2,3)$).}
    \label{clean_case2_mapa}
\end{figure}

\begin{figure}
    \centering
    \includegraphics[width=\textwidth]{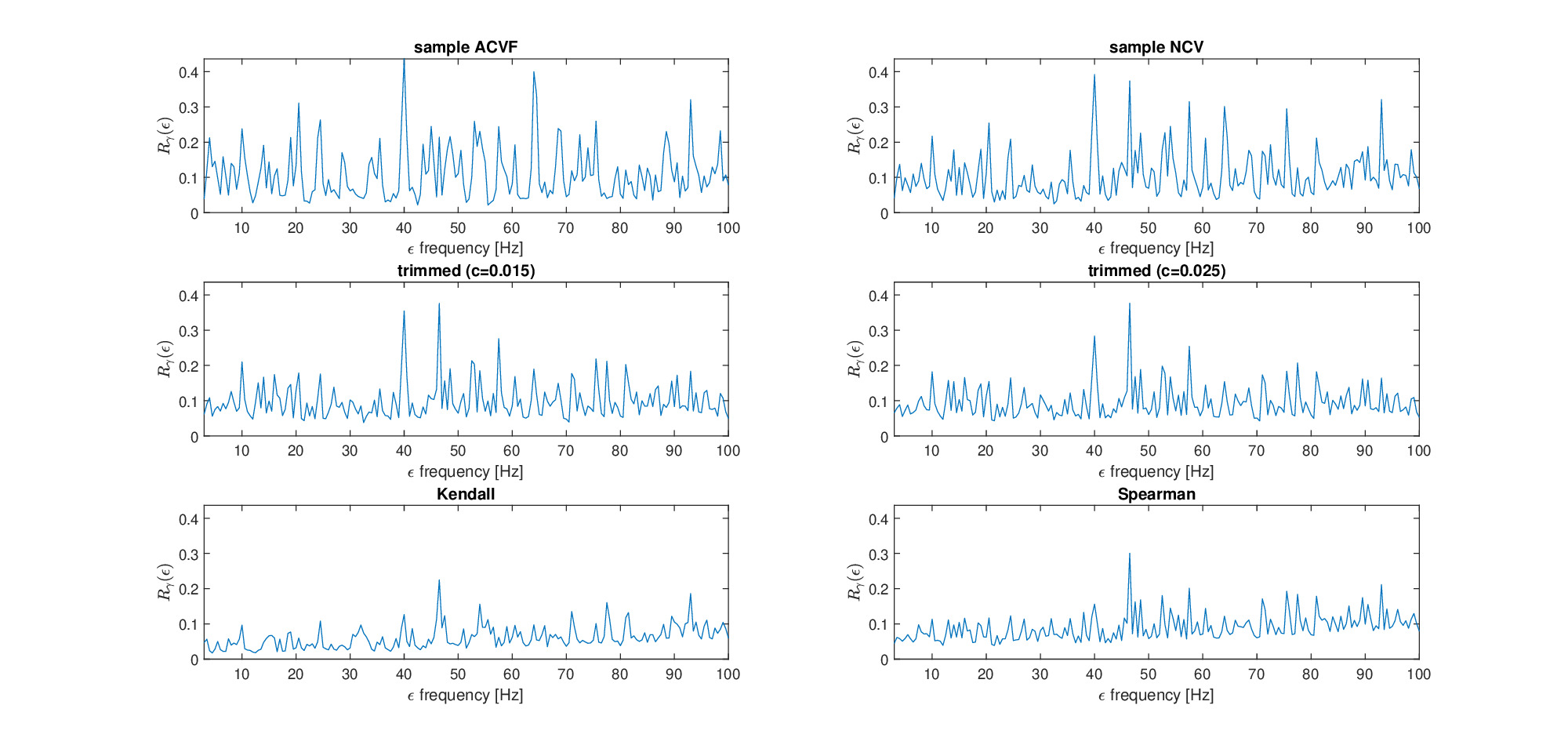}
    \caption{Statistics $R^\prime_\gamma(\epsilon)$ for Case 2 without cyclic impulses ($\{Z_t\} \sim \mathcal{T}(2,3)$).}
    \label{clean_case2_amp}
\end{figure}

\begin{figure}
    \centering
    \includegraphics[width=\textwidth]{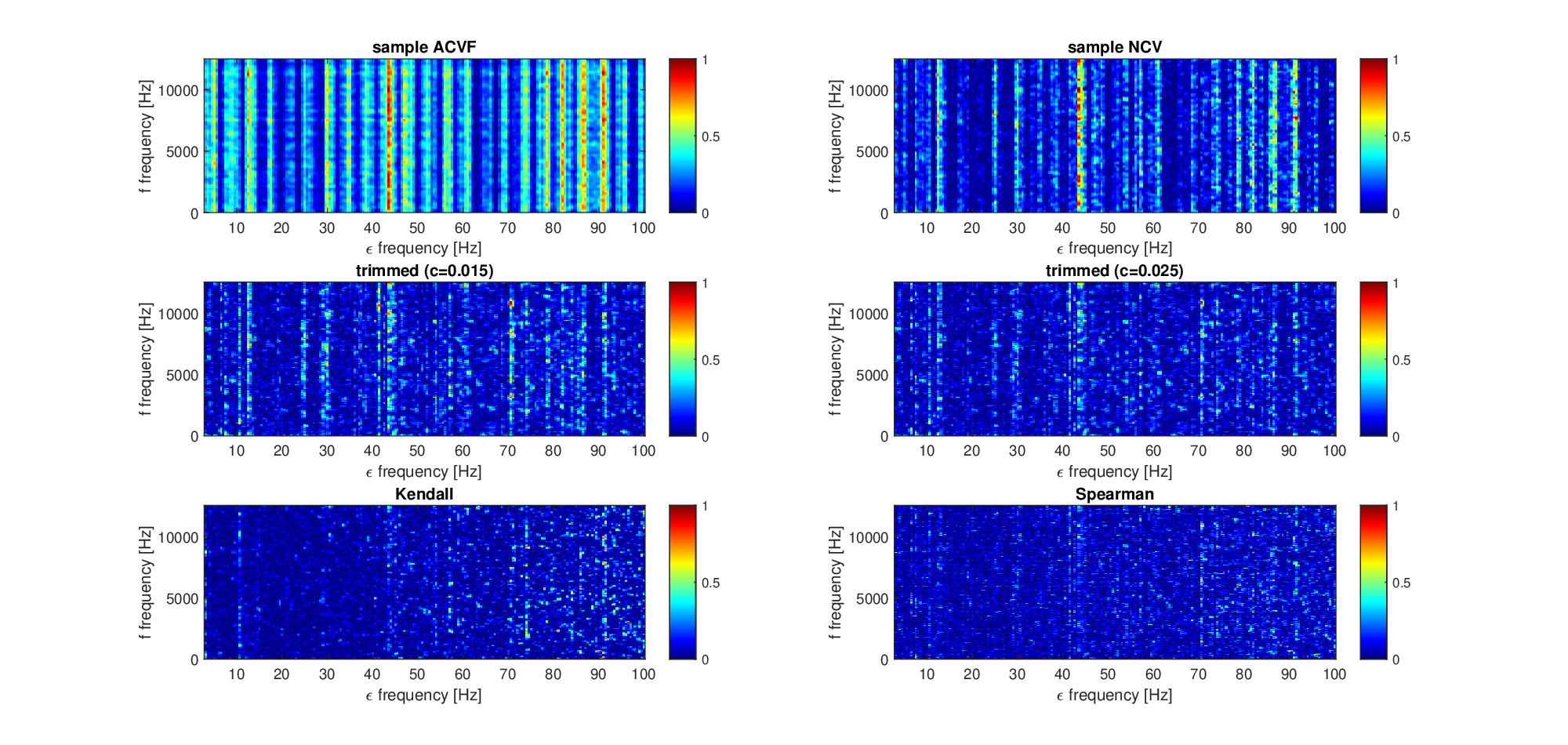}
    \caption{Spectral coherence maps $|\gamma(f,\epsilon)|^2$ for Case 3 without cyclic impulses ($\{Z_t\} \sim \mathcal{S}(1.7,3)$).}
    \label{clean_case3_mapa}
\end{figure}

\begin{figure}
    \centering
    \includegraphics[width=\textwidth]{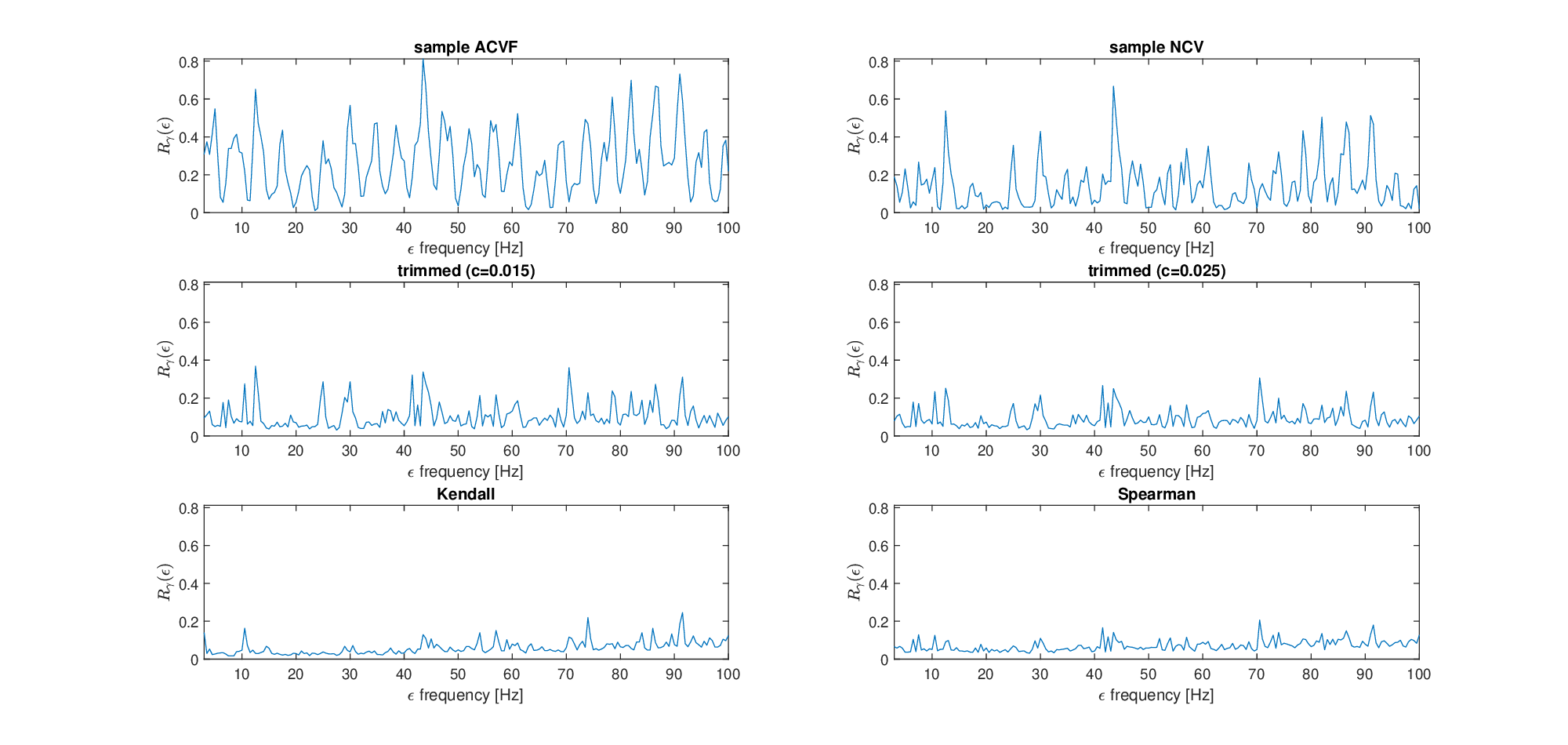}
    \caption{Statistics $R^\prime_\gamma(\epsilon)$ for Case 3 without cyclic impulses ($\{Z_t\} \sim \mathcal{S}(1.7,3)$).}
    \label{clean_case3_amp}
\end{figure}

The maps obtained for Case 1 without cyclic impulses are presented in Fig. \ref{clean_case1_mapa}. The corresponding plots of $R^\prime_\gamma(\epsilon)$ statistic for each map are shown in Fig. \ref{clean_case1_amp}. For Case 2 without cyclic impulses, the analogous results are illustrated in Figs. \ref{clean_case2_mapa} and \ref{clean_case2_amp}. The maps and $R^\prime_\gamma(\epsilon)$ statistics for Case 3 without cyclic impulses are presented in Figs. \ref{clean_case3_mapa} and \ref{clean_case3_amp}. In all cases, the periodicity was correctly not detected. However, one can see that the proposed robust maps (in particular for Kendall and Spearman estimators) are much less disturbed than the classical one for sample ACVF.

\begin{figure}
    \centering
    \includegraphics[width=\textwidth]{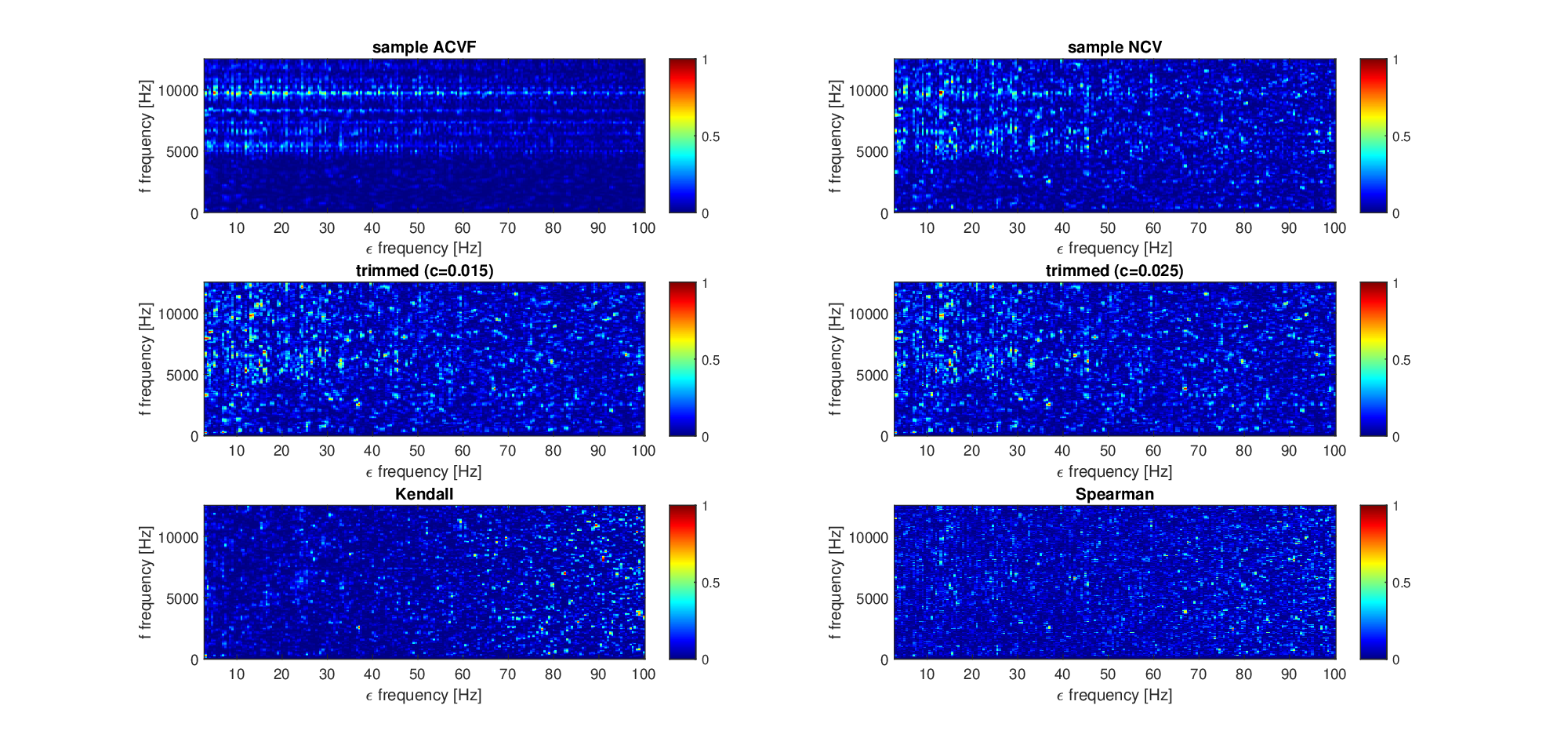}
    \caption{Spectral coherence maps $|\gamma(f,\epsilon)|^2$ for the Dataset 1 (real vibration signal without periodic impulses).}
    \label{ff_dataset1_map}
\end{figure}

\begin{figure}
    \centering
    \includegraphics[width=\textwidth]{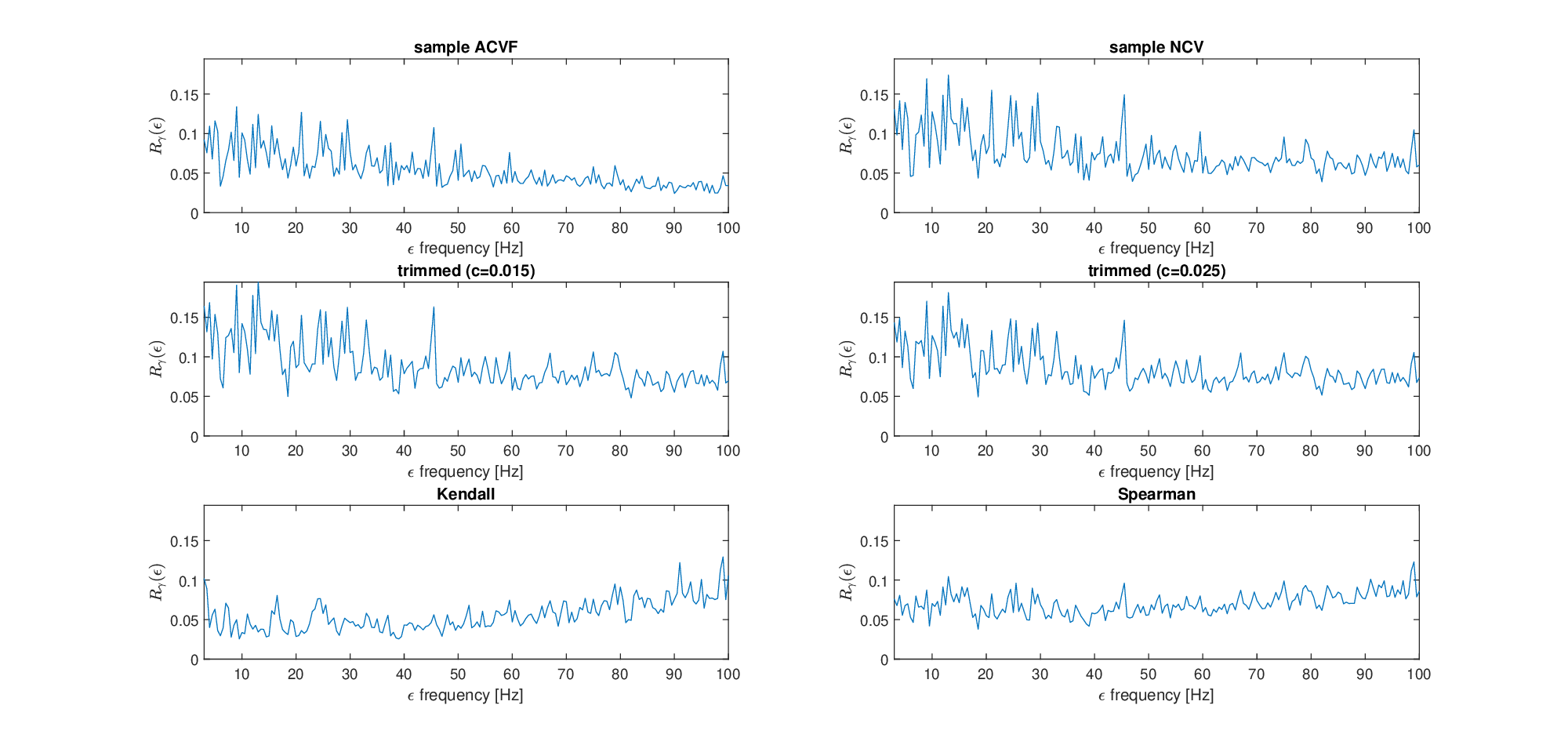}
    \caption{Statistics $R^\prime_\gamma(\epsilon)$ for the Dataset 1 (real vibration signal without periodic impulses).}
    \label{ff_dataset1_amp}
\end{figure}

Analogously, let us now consider the Dataset 1, i.e. the real vibration signal from a healthy machine introduced in Section \ref{real}. In Fig. \ref{ff_dataset1_map}, we present the maps and $R^\prime_\gamma(\epsilon)$ defined in Eq. (\ref{rprim}) obtained for this dataset. Again, as expected, no periodicity was detected. However, let us note that the sample ACVF-based map has visibly larger values for $f>5000$ Hz than for lower $f$ (similarly as for Dataset 2, see Section \ref{real} and Fig. \ref{ff_dataset2_map}). Such behavior is clearly not desired and can be considered a disadvantage of this method for the analyzed case. In Fig. \ref{ff_dataset1_amp}, we present the corresponding $R^\prime_\gamma(\epsilon)$ statistics. Here, both trimmed estimators exhibit similar behavior as the sample ACVF, however again both Kendall and Spearman estimators seem to yield most homogeneous results. 

\end{document}